\documentclass[floatfix,aps,pre,showpacs]{revtex4}
\usepackage{graphicx}
\usepackage{amsmath}
\usepackage{amsfonts}
\usepackage{amssymb}
\begin{document}

\title{Dynamic mean-field and cavity methods for diluted Ising systems}

\author{Erik Aurell}
\affiliation{Department of Computational Biology, AlbaNova University Centre, 106 91 Stockholm, Sweden}
\altaffiliation{Department of Information and Computer Science, Aalto University, Finland}
\altaffiliation{ACCESS Linnaeus Centre, KTH - Royal Institute of Technology, Stockholm, Sweden}
\author{Hamed Mahmoudi}
\affiliation{Department of Information and Computer Science, Aalto University, Finland}

\date{\today}

\begin{abstract}
We compare dynamic mean-field and dynamic cavity as methods
to describe the stationary states of dilute kinetic Ising models.
We compute dynamic mean-field theory by expanding in interaction
strength to third order, and compare to the exact dynamic 
mean-field theory for fully asymmetric networks.
We show that in diluted networks the dynamic cavity method  
generally predicts magnetizations of individual spins
better than both first order (``naive'') and second order
(``TAP'') dynamic mean field theory.   
\end{abstract}


\maketitle 
\section{Introduction}
(Classical) statistical mechanical systems in equilibrium are described by the Gibbs measure, which
connects the propensity of a system to move between two states taken in isolation
(the energy differences between these two states) to the probability of finding the
system in one of the states, when all states are available. This relation is
normally used to find equilibrium statistics of a system (magnetizations, correlation functions
etc.) by sampling a dynamics for which the Gibbs measure in a stationary state.
A Markov Chain Monte Carlo (MCMC) method, such as Glauber dynamics for Ising systems,
which we will review briefly below in Sec.~\ref{sec:model}, can work if the sample
average converges quickly enough to Gibbs measure, and if the quantity to be measured
has wide support in phase space. Well-known scenarios where this is case are spin systems in
the high-temperature phase and when measuring \textit{e.g.} total magnetization. In the low-temperature
phase relaxation time to the Gibbs distribution can be very long. On the other hand, if the
quantity to be measured is \textit{e.g.} the magnetization of a single spin, then MCMC in
a large system is slow for the trivial reason that one needs to sweep through all the spins
while just being interested in the changes in and average over one of them.
If the interactions are weak, marginal probability distributions can be computed
perturbatively in mean-field theory~\cite{TAP,Parisi-book,Tanaka2000}, which give closed equations
for \textit{e.g.} single-spin magnetizations. For dilute systems, where every spin is not connected
to most other spins, very powerful
\textit{message-passing} methods have been developed by physicists, information theorists and computer scientists over
the last two decades to compute marginals of Gibbs distributions quickly and accurately~\cite{YedidiaFreemanWeiss03,MezardMontanari09}.
While these \textit{cavity equations} cannot (in their simplest form) deal with the complex phases of random
spin systems at low temperature, in suitable scenarios they are much more accurate than 
mean-field theory, and they greatly improve on MCMC for single site magnetization and other local properties
by substituting a cumbersome sampling by a direct deterministic computation. 
The cavity method has found to have many
technological as well as fundamental applications~\cite{Pearl88,Pearl,opper2001adaptive,Kschischang01,MezardMontanari09}.

The situation is very different for out-of-equilibrium systems, in itself is an
extremely broad term covering everything from macroscopic hydrodynamics (turbulence)~\cite{Frisch-book}
and physical and chemical kinetics~\cite{vanKampen-book}
to interdisciplinary applications of statistical physics to neuroscience, 
population biology, and other fields~\cite{hopfield,BlytheMcKane}.
We here consider the model systems obtained when
generalizing the MCMC rules of Ising spin systems (Glauber dynamics) from 
the equilibrium case (symmetric interactions) to a non-equilibrium
case (non-symmetric interactions). Such ``kinetic Ising'' models
are only conceptual -- but tractable -- models of real spin systems
driven out of equilibrium, and have mainly been studied with 
applications to neuroscience in 
mind~\cite{derrida1987exactly,HertzGrinsteinSolla86,HertzGrinsteinSolla87,sompol1}. 
From the mathematical point of view, they are specific
examples of Markov chains which do not obey detailed balance conditions.
In contrast to equilibrium systems, there is hence no simple expression
for a stationary state akin to a Gibbs measure, but such a state, when it
exists, is a (complicated) function of all the details of the model.
On the other hand, MCMC works as well in such systems as for standard equilibrium 
Ising models, and mean-field theory have been developed 
up to second order in the interaction strength~\cite{KappenSpanjers2000}.
This leaves open the case of dilute kinetic Ising models, where in
the equilibrium case cavity methods would be preferable. 
A dynamic cavity method has only very recently 
developed for majority dynamics\cite{KanoriaMontanari} and
Glauber dynamics~\cite{NeriBolle} and 
was investigated by us for parallel and
sequential update schemes in~\cite{AurellMahmoudi2011}.

The dynamic cavity method as outlined in~\cite{AurellMahmoudi2011} comprises
first an ansatz on probability distributions, similar to standard cavity, 
then the Belief Propagation ansatz that cavity distributions factorize, 
and then also a further assumption of Markovianity. As discussed in \cite{AurellMahmoudi-2011b},
the second assumption is exact for fully asymmetric models with the parallel update rule.
In a more general case, where either the interaction matrix
has both a symmetric and an asymmetric component, or where the update rule is different,
it is however but an approximation. 
The numerical results of~\cite{AurellMahmoudi2011}, which showed that for some such mixed instances the
dynamic cavity is quite accurate, were somewhat unexpected. The main motivation for the present
paper is therefore to show more systematically in what parameter ranges dynamic
cavity converges (for these models), where it is accurate compared to MCMC, and
to compare its predictions to mean field theory. We will show that for dilute kinetic
Ising models, dynamic cavity works also for the magnetizations of individual spins,
and is considerably more accurate than mean-field calculations of the same quantities.

Kinetic Ising models have been studied by other approaches, and we outline them
briefly here.
When the discreteness of states is relaxed to a spherical Ising model.
Sompolinsky and Zippelius developed a Langevin equation formalism~\cite{sompol2},
later extended by Crisanti and Sompolinsky to the non-equilibrium case~\cite{CrisantiSompolinsky1987},
where several phases of these (dynamical) models are outlined. Although pioneering,
predicting magnetizations of individual spins is out of scope of such methods, as the sphericity 
approximation has been made.
The dynamic replica theory (DRT) has been applied to kinetic Ising models~\cite{coolen1},
which, by the nature of replica theory, however only applies to averages over
ensembles of models. 
Sommers developed a path integral formulation for the Glauber dynamics~\cite{sommers},
which was at that time only investigated approximately. 
As an alternative approach to path integral formulation, generating functional analysis
was developed to study non equilibrium statistical mechanics of disordered systems~\cite{coolen2}.
It was shown by Neri and Boll\'e in~\cite{NeriBolle} that
at least in some cases, a dynamic cavity analysis explicitly averaged over 
a random ensemble recovers the results of  generating functional analysis.
Recently, Hertz and 
Roudi~\cite{HertzRoudi2011,RoudiHertz2011} used generating functional analysis to derive mean-field theories,
for infinite-range spin glass models.
To compare the accuracy of the predictions
of single-site magnetizations by the dynamic mean-field formula of~\cite{HertzRoudi2011}
to the dynamic cavity for dilute mixed models was one further motivation for this work.

The paper is organized as follows. In section~\ref{sec:model} we 
describe the Glauber dynamics for spin glasses, the model which we will study. 
In Section~\ref{sec:infinite_range} we discuss two approaches to a dynamic
version of the TAP corrections to first-order mean-field
theory~\cite{KappenSpanjers2000,HertzRoudi2011,RoudiHertz2011,Mezard2011}, while in
Section~\ref{sec:BP} we derive dynamic cavity equations for diluted networks in parallel update. 
This derivation should be seen as an alternative and (we hope)
clearer alternative to~\cite{NeriBolle} and our earlier contribution~\cite{AurellMahmoudi2011}.
The main new results of this paper, on the convergence phase of dynamic cavity and on
a comparison between the predictions of dynamic cavity and mean-field theory to MCMC are presented
in Section~\ref{sec:results}. In Section~(\ref{sec:conclusion}) we conclude and discuss possible
application areas of dynamic cavity.

\section{The parallel spin update scheme in dilute kinetic Ising models}
\label{sec:model}

The asymmetric dilute Ising model is defined over 
a set of $N$ binary variables $\vec{\sigma}=\{\sigma_1,\ldots,\sigma_N\}$,
and an asymmetric graph $G=(V,E)$ where
$V$ is a set of $N$ vertices, and $E$ is a set of directed edges.
To each vertex $v_i$ is associated a binary variable $\sigma_i$.
The graphs $G$ are taken from random graph ensembles with
bounded average connectivity. Following the parametrization of~\cite{coolen2}
we introduce a connectivity matrix $c_{ij}$, where $c_{ij}=1$ if there is
a link from vertex $i$ to vertex $j$, $c_{ij}=0$ otherwise, and matrix elements
$c_{ij}$ and $c_{kl}$ are independent unless $\{kl\}=\{ji\}$.  
The random graph is then specified by marginal (one-link) distributions
\begin{equation}
p(c_{ij}) = \frac{c}{N}\delta_{1,c_{ij}} + (1 - \frac{c}{N})\, \delta_{0,c_{ij}}\,\,\,.
\label{eq:connec}
 \end{equation}   
and conditional distributions
\begin{equation}
p(c_{ij}\,|\,c_{ji}) = \epsilon \delta_{c_{ij},c_{ji}} + (1-\epsilon) \, p(c_{ij})\,\,\,.
\label{eq:asym}
\end{equation} 
where $i,j\in\{1,...,N\}$ and $i<j$. In this model the average degree distribution is given by $c$,
and the asymmetry is controlled by $\epsilon\in[\,0,1\,]$.
The two extreme values of $\epsilon$ give respectively a fully asymmetric network 
($\epsilon=0$), where the probabilities of having two directed links
between pairs of variables are uncorrelated, and the symmetric network ($\epsilon=1$)
where the two links $i\to j$ and $j\to i$ are present or absent together. 
The parameter set is completed by a (real-valued) interaction matrix
$J_{ij}$. 
Additional assumptions on the $J_{ij}$, \textit{i.e.} smallness or that
they are random with suitable distribution are stated when used. However, for
concreteness the reader may in much of this paper think of $J_{ij}$ 
to be independent identically distributed random variables with 
zero mean and variance $\frac{1}{c}$ (Gaussian or binary) such that for the fully
connected networks $(c=N)$, the interactions scale as the 
Sherrington-Kirkpatrick model~\cite{SherringtonKirkpatrick75}. 

The interactions among spins determine the dynamics of system. 
In the parallel update scheme, which will be considered here,
at each (discrete) time,
all spins are updated according to the Glauber rule
\begin{eqnarray}
\sigma_i(t+1) = \left\{
\begin{array}{c cl}
+1 &{\rm with \,\,probability} \,\,\,\, \{1+\exp(-2\beta\,h_i(t+1))\}^{-1}&\\
-1  &{\rm with \,\,probability} \,\,\,\, \{1+\exp(2\beta\,h_i(t+1))\}^{-1}& 
\end{array}
\right.
\label{eq:dynamic}
\end{eqnarray}
where $h_i(t)$ is the effective field acting on spin $i$ at time step $t$
\begin{equation}
h_i(t) = \sum_{j\in\partial i} \, J_{ji}\,\sigma_j(t-1)+ \theta_i(t) \,\,\,\,\, . 
\label{eq:eff_field}
\end{equation}
and the parameter $\beta$, analogous to inverse temperature, is
a measure of the overall strength of the interactions.
The notation $j \in \partial i$ in (\ref{eq:dynamic}) and
(\ref{eq:eff_field})
indicates all vertices having a direct
links to node $i$ and $\theta_i$ is the
(possibly time-dependent) external field acting on spin $i$.
In this paper we will adhere to the convention that the interaction
indices are written in the same order as the temporal order of
the interacting spins. Hence we have $J_{ij}\sigma_i(s)\sigma_j(s+1)$
and $J_{ji}\sigma_j(s)\sigma_i(s+1)$.

The joint probability distribution over all the spin histories
$p(\vec{\sigma}(0),\ldots,\vec{\sigma}(t))$ has in principle the following simple
Markov form
\begin{equation}
P(\vec{\sigma}(0)\ldots,\vec{\sigma}(t)) = \prod_{s=1}^t \,W[\,\vec{\sigma}(s)\,|\,\vec{h}(s)]\,p(\vec{\sigma}(0))
\label{eq:prob_dyn}
\end{equation} 
where $W$ is the appropriate transition matrix describing dynamics and 
updates. 
If we would have a fully understanding of joint probability distribution defined in (\ref{eq:prob_dyn}) we 
could compute
time dependent macroscopic quantities such as magnetization and correlations.
The evolution of a a single spin is (trivially) defined by 
summing over the histories of all spins except one
\begin{equation}
P_i(\sigma_i(0),\ldots,\sigma_i(t)) = \sum_{\vec{\sigma}_{\setminus i}(0),\ldots,\vec{\sigma}_{\setminus i}(t)} \, P(\vec{\sigma}(0),\ldots,\vec{\sigma}(t))
\label{eq:marginal}
\end{equation}
which can be further marginalized to the probability of one spin at one time
\begin{equation}
p_i(\sigma_i(s)) = \sum_{\sigma_{i}(0),\ldots,\sigma_i(s-1),\sigma_i(s+1),\ldots,\sigma_{ i}(t)} \, P_i(\sigma_i(0),\ldots,\sigma_i(t)) 
\label{eq:marginal}
\end{equation}
and similarly  for pairwise joint probability of the histories of two spins
$P_{ij}(\sigma_i(0),\ldots\sigma_i(t),\sigma_j(0),\ldots,\sigma_j(t))$ and
$p_{ij}(\sigma_i(s),\sigma_j(s'))$.
Consequently, the time evolution of single site magnetization can be obtained from Eq(\ref{eq:marginal}) as
\begin{equation}
m_i(t) = \sum_{\sigma_i(t)} \sigma_i(t) \,p_i(\sigma_i(t))
\label{eq:mag}
\end{equation}
and similarly the correlation functions
\begin{equation}
c_{ij}(s,t) = \sum_{\sigma_i(s),\sigma_j(t)} \sigma_i(s) \sigma_j(t) p_{ij}(\sigma_i(s),\sigma_j(t))
\label{eq:corr}
\end{equation}
Substituting Eq(\ref{eq:mag}) into dynamics defined in (\ref{eq:prob_dyn}) we have 
\begin{equation}
m_i(t) = \left<\tanh(\sum_{j\in\partial i} J_{ji} \sigma_j(t-1) + \theta_i(t))\right>
\label{eq:exact-magnetization}
\end{equation}
where brackets are average with respect to trajectory history.
Equation~(\ref{eq:exact-magnetization}) is exact for the time-dependent magnetization.
It is not directly practical, since the marginal over one spin at one
time (the magnetization) depends on the joint distribution of all the spins
influencing it at the previous time, but as we will see 
in Section~\ref{sec:information-geometry} it can be used as
a starting point of a perturbative calculation. 


\section{Mean-field theories, TAP, and the expansion in small interactions}
\label{sec:infinite_range}

The mean field theory of spin glass systems started with the 
Sherrington Kirkpatrick (SK) model~\cite{SherringtonKirkpatrick75}. In this model all spins
interact with all other spins (infinite-range couplings), which motivates the simplest mean field
or ``naive mean-field'' approximation 
$m_i=\tanh\beta\left(\sum_{j}J_{ji}m_j+\theta_i\right)$. 
Shortly afterwards a more accurate mean field theory (TAP) was 
introduced by introducing Onsager reaction for the SK model. This corrects 
$m_j$ inside the $\tanh$ to $m_j-\beta J_{ij}m_i(1-m_j)^2$ where $J_{ij}m_i$ is the field from
spin $i$ on spin $j$ and where $\chi_{jj}=\beta(1-m_j^2)$ can be interpreted as a local susceptibility at spin $j$~\cite{TAP}.
Since in equilibrium Ising $J_{ij}=J_{ji}$ the TAP equilibrium mean field theory is hence 
$m_i=\tanh\left(\beta\sum_{j}J_{ji}m_j+\beta\theta_i-\beta^2m_i\sum_j J_{ji}^2(1-m_j^2)\right)$. 
As stated in~\cite{TAP} these results can be derived from the cavity approach. 
These can also be derived by observing that in equilibrium 
a susceptibility is related to a correlation by
fluctuation-dissipation, and the appropriate correlation was computed by 
a perturbative argument~\cite{TAP}. For a later approach by 
field-theoretical methods, expanding a functional determinant
describing the fluctuations around a mean-field stationary point
of an action, see \textit{e.g.}~\cite{kholodenko1990onsager}, and references therein.

In equilibrium Ising systems the naive mean-field and the TAP approximations can further be computed
by expanding the Boltzmann distribution in the interaction strength~\cite{plefka1982convergence} . 
To first and second order in interactions this result agree with naive mean-field
and TAP. 

Recently, a dynamic version of TAP
has been derived by 
Hertz and Roudi
~\cite{HertzRoudi2011,RoudiHertz2011}
by a field-theoretical argument, and we show here in Section~\ref{sec:information-geometry}
below that this also follows from Information Geometry, essentially a systematic
expansion in interaction strength.
For completeness, we will also show that the same dynamic version of TAP 
follows from the ``exact mean-field theory'' of M\'ezard and 
Sakellariou~\cite{Mezard2011}, as already pointed out in~\cite{sakellariou2011effect}.

Outside equilbrium fluctuation-dissipation does not hold. Conceptually one could therefore
say that ``dynamic TAP'' as such is undefined, or, alternatively, that a proper generalization of TAP
to a non-equilibrium system should be based on fluctuation relations generalizing 
fluctuation-dissipation theorems~\cite{marconi2008fluctuation} (a task we have not attempted to carry out).
In this paper we take however a more pragmatic approach, and understand ``dynamic TAP'' to be the formulae
derived in~\cite{HertzRoudi2011,RoudiHertz2011}.

Before turning to the technical discussion, let us note that since
mean-field and TAP have obvious computational advantages, these
theories have been applied in much wider settings than in which
they have been derived, particularly in neuroscience. For a recent review, 
see~\cite{hertz2011ising} and references therein.

\subsection{Fully asymmetric networks: a reduced theory}
\label{sec:fully_asymmetric}
In this section we recall the theory in~\cite{Mezard2011}, with a view to
compute the expansion in small interactions to third order.
We start by rewriting the exact equation (\ref{eq:exact-magnetization}) in the
following explicit form:
\begin{equation}
m_i(t) = \sum_{\sigma_i(t),\sigma_{\partial i}(t-1)}\,\, p(\sigma_{\partial i}(t-1))\,\,
\sigma_i(t)\,\, \frac{e^{\beta \sigma_i(t)\left(\sum_{j\in\partial i}J_{ji}\sigma_j(t-1)+\theta_i(t)\right)}}{2\cosh\left(\beta(\sum_{j\in\partial i}J_{ji}\sigma_j(t-1)+\theta_i(t))\right)}
\label{eq:exact-explicit-magnetizations}
\end{equation}
where $\sigma_{\partial i}$ is the collection of spins neighboring $i$ with $c_{ji}\neq 0$
and $p(\sigma_{\partial i}(t-1))$ is the corresponding joint probability distribution.
In a fully asymmetric network, when an interaction coefficient $J_{ji}$ in above is non-zero,
then the opposite $J_{ij}$ is zero. Each of the spins $\sigma_j(t-1)$ on the right hand side therefore
does not depend directly on spin $i$ on yet one time step before, \textit{i.e.} on $\sigma_{i}(t-2)$.
Furthermore, the distribution of each of the $\sigma_j(t-1)$ will in turn depend on distributions
of other $\sigma_k(t-2)$, but the distribution of these $\sigma_k(t-2)$ do not depend on the $\sigma_j(t-1)$.
If there are no short paths in the interaction graph between any pairs of spins 
$\sigma_j$ on the right hand side of (\ref{eq:exact-explicit-magnetizations})
except through the cavity spin $\sigma_i(t)$, or if such paths are unimportant,
then the spins $\sigma_j(t-1)$ will be independent in an asymmetric network.
and the effective field $h_i(t)=\theta_i(t)+\sum_{j\in\partial i}J_{ji}\sigma_j(t-1)$ acting on $\sigma_{i}(t)$
will be the sum of independent random variables.

When the number of interacting spins is large the distribution of $h_i(t)$ follows from the
central limit theorem 
\begin{equation}
p(h_i(t)) = \frac{1}{\sqrt{2\pi V_i(t)}} \exp{\left[-\frac{(h_i(t)-\langle h_i(t)\rangle)^2}{2 V_i(t)}\right]}
\end{equation} 
where $\langle h_i(t)\rangle = \theta_i(t)+\sum_{j\in\partial i}J_{ji} m_j(t-1)$ and $V_i(t) = \langle h_i(t)\rangle^2 - \langle h_i^2(t)\rangle$.
We note that to arrive at this result, first the thermodynamic limit ($N\to\infty$) is taken
at given connectivity $c$ (so that the terms $J_{ji}m_j(t-1)$ are independent), and then $c$ is taken large
(so that there are many of them). In general $V_i(t)$ is defined as 
\begin{equation}
V_i(t) = \sum_{j\in \partial i, k\in \partial i} J_{ji} J_{ki} \left[\langle\sigma_j(t-1) \sigma_k(t-1)\rangle-m_j(t-1) m_k(t-1) \right]
\end{equation}
When the additional assumption that the interaction coefficients $J_{ji}$ are random, independent and evenly distributed 
and small the sum is dominated by the diagonal terms \textit{i.e.}
\begin{equation}
V_i(t) = \sum_{j\in\partial i} J_{ji}^2 \left(1-m_j^2(t-1)\right)
\end{equation}
This gives the ``exact mean-field'' theory of~\cite{Mezard2011}:
\begin{equation}
m_i(t) = \int\, Dx \,\,\tanh\left[\beta \left(\theta_i(t) + \sum_{j} J_{ji} m_j(t-1) + x\sqrt{\sum_{j}J_{ji}^2(1-m_j(t-1)^2)}\right)\right]
\label{eq:asym_mf}
\end{equation}
with the Gaussian measure $Dx = \frac{dx}{\sqrt{2\pi}}e^{-\frac{x^2}{2}}$.\
Equation~(\ref{eq:asym_mf}) can be iterated starting from some initial condition to get
all magnetizations at any time, and is exact when the
assumptions hold \textit{i.e} when the network is fully asymmetric, when the cavity assumptions
hold, when any spin is influenced by a large number of other spins, and when the interactions
are random, independent, evenly distributed and small. 

To expand~(\ref{eq:asym_mf}) in small interactions we 
introduce $c_i(t)\equiv \sqrt{\sum_{j}J_{ji}^2(1-m_j(t-1)^2)}$
and take all these quantities of order $\epsilon$. 
We have
\begin{eqnarray}
\tanh\left[\beta \left(\langle h_i(t)\rangle+c_i(t)x\right)\right] &=& 
\tanh\left[\beta \langle h_i(t)\rangle\right] + xc_i(t)\beta(1-\tanh^2\left[\beta \langle h_i(t)\rangle \right])
\\ \nonumber -
&& \quad x^2c_i^2\beta^2\tanh\left[\beta \langle h_i(t)\rangle\right](1-\tanh^2\left[\beta \langle h_i(t)\rangle\right])+ {\cal O}(\epsilon^3)
\end{eqnarray}
where every odd term in this expansion will give zero when integrated against a Gaussian measure. Therefore we have
\begin{equation}
m_i(t) = \tanh\left[\beta \langle h_i(t)\rangle\right] - \beta^2\tanh\left[\beta \langle h_i(t)\rangle\right](1-\tanh^2\left[\beta \langle h_i(t)\rangle\right])c_i^2(t) +  {\cal O}(\epsilon^4)
\label{eq:tanh-fourth}
\end{equation}
We would like to write the right hand side of (\ref{eq:tanh-fourth}) as  
$\tanh\left[\beta \left(\langle h_i(t)\rangle+\Delta_i(t)\right)\right] + {\cal O}(\epsilon^4)$. A comparison shows that this is possible 
setting $\Delta_i(t)= \beta\tanh\left[\beta \langle h_i(t)\rangle\right]c_i^2(t)$. We therefore have to \textit{fourth} order the
following functional expression
\begin{equation}
m_i(t) = \tanh\left[\beta\left(\langle h_i(t)\rangle - \beta\tanh\left[\beta \langle h_i(t)\rangle\right]c_i^2(t)\right)\right] +  {\cal O}(\epsilon^4)
\label{eq:tanh-functional}
\end{equation}
To first order in $\epsilon$ the solution is
\begin{equation}
m_i(t) = \tanh\left[\beta\left(\sum_{j\in\partial i} J_{ji} m_j(t-1) + \theta_i(t)\right)\right]+  {\cal O}(\epsilon^2)
\label{eq:naive-mean-field}
\end{equation}
which is the ``dynamic naive mean-field''.
Inserting this back in (\ref{eq:tanh-functional}) we have ``dynamic TAP'' of~\cite{HertzRoudi2011,RoudiHertz2011}
\begin{equation}
m_i(t) = \tanh\left[\beta\left(\sum_{j\in\partial i} J_{ji} m_j(t-1) + \theta_i(t)\right)-\beta^2 m_i(t)
          \sum_{j\in\partial i} J_{ji}^2 (1-m_j(t-1)^2\right]
+  {\cal O}(\epsilon^4)
\label{eq:tap_dynamic}
\end{equation}
The last term inside the $\tanh$ is of order $\epsilon^2$ and a form analogous to the Onsager back-reaction term;
there is no third order correction in $\epsilon$ in this theory.


\subsection{The Information Geometry viewpoint}
\label{sec:information-geometry}
We will now derive the analogues of (\ref{eq:naive-mean-field}),  (\ref{eq:tap_dynamic})
and a third order term by following the approach of Information Geometry~\cite{Tanaka2000,KappenSpanjers2000,AmariIkedaShimokawa2001}.
Let $\vec\sigma(0),\ldots,\vec\sigma(t)$ be the time history of a collection of
spins. We assume that these spins have been generated by a kinetic Ising model
with parallel updates and (possibly) time-dependent external fields. The joint probability
of the history of all the spins is then
\begin{equation}
P(\vec\sigma(0),\ldots,\vec\sigma(T)|\vec\theta(0),\ldots,\vec\theta(T),\{J_{ij}\}) =
\prod_{t=1}^T\prod_i \exp(\sigma_i(t)h_i(t))/2\cosh(h_i(t)), \,\, h_i(t)=\theta_i(t)+\sum_j J_{ji}\sigma_j(t-1).
\label{eq:Glauber-definition}
\end{equation}
In Information Geometry the space of these model is considered as a manifold
with coordinates being the (many) parameters $\vec\theta(0),\ldots,\vec\theta(T),\{J_{ij}\}$.
A sub-manifold is the family of independent models 
\begin{equation}
P^{\hbox{ind}}(\vec\sigma(0),\ldots,\vec\sigma(T)|\vec\theta^{\hbox{ind}}(0),\ldots,\vec\theta^{\hbox{ind}}(T)) =
\prod_{t=1}^T\prod_i \exp(\sigma_i(t)h_i(t))/2\cosh(h_i(t)), \,\, h_i(t)=\theta^{\hbox{ind}}_i(t).
\label{eq:independent-model-definition}
\end{equation}
A mean-field approximation is defined as the
the independent model with the same magnetizations as the full model~\cite{AmariIkedaShimokawa2001}. 
For our case it is easily seen that $m_i(t)=\tanh(\theta_i^{\hbox{ind}}(t))$ is the variational equation
with respect to parameter $\theta_i^{\hbox{ind}}(t)$ of the 
Kullback-Leibler divergence $D_{-1}\left[p | p^{\rm{ind}}\right]=\sum p\ln\frac{p}{p^{\rm{ind}}}$.
Therefore, the mean field approximation in Information Geometry can also be seen
as the independent model with the least Kullback-Leibler divergence from 
the full model~\cite{Tanaka2000,KappenSpanjers2000,AmariIkedaShimokawa2001}.

Following the approach of~\cite{KappenSpanjers2000} we take the interaction parameters ($J_{ij}$) as
small parameters (of order $\epsilon$), and assume that the differences
$\Delta\theta_i(t)=\theta_i(t)-\theta_i^{\hbox{ind}}(t)$ can be expanded in $\epsilon$:
\begin{equation}
\Delta\theta_i(t)=\epsilon \Delta^{(1)}_i(s) + \epsilon^2  \Delta^{(2)}_i(s) + \ldots
\label{eq:delta-expansion}
\end{equation}
We can then write
in analogy with Eq.3.2 of~\cite{KappenSpanjers2000} 
\begin{eqnarray}
0&=&m_i(t)-m_i^{\hbox{ind}}(t) = 
  \epsilon \sum_{i,s} \frac{\partial m_i(t)}{\partial\theta_i(s)}|_{\hbox{ind}}\Delta^{(1)}_i(s)
+ \epsilon \sum_{j,k} \frac{\partial m_i(t)}{\partial J_{kl}}|_{\hbox{ind}} J_{kl}  \nonumber \\
&&+ \epsilon^2 \sum_{i,s} \frac{\partial m_i(t)}{\partial\theta_i(s)}|_{\hbox{ind}}\Delta^{(2)}_i(s) 
+\frac{\epsilon^2}{2}\sum_{JK} \frac{\partial^2 m_i(t)}{\partial \Theta_J\partial\Theta_K}|_{\hbox{ind}}\Delta^{(1)}\Theta_J\Delta^{(1)}\Theta_K + {\cal O}(\epsilon^3)
\label{eq:Kappen-Spanjers-3.2}
\end{eqnarray}
Here $\Theta_J$ stands for the set of all interacting couplings and external fields and $J$ runs over relevant indices. The subscript indicates that all
derivatives are evaluated at the independent model, and the left-hand side is zero because this
is the variational equation.
In the last term the sum goes over all the parameters 
labeled $J,K$ and the parameter increments are the first order terms
$\Delta^{(1)}_i(s)$ and $J_{kl}$; on third and higher orders mixed terms
of $\Delta^{(1)}_i(s)$  and $\Delta^{(2)}_i(s)$ will appear. A calculation presented in Appendix gives the results
\begin{eqnarray}
\Delta^{(1)}_i(t) &=& - \sum_{j} J_{ji}m_j(t-1) \label{eq:delta-first-order} \\
\Delta^{(2)}_i(t)    &=& m_i(t)\sum_k J_{ki}^2 (1-m_k^2(t-1)) \label{eq:delta-second-order} \\
\Delta^{(3)}_i(t) &=& - \sum_k (1-m_k^2(t-1)) J_{ki} \Delta^{(2)}_k(t-1) 
\end{eqnarray}
Equation~(\ref{eq:delta-expansion}) together with the variational equation
can be re-written
\begin{equation}
\tanh^{-1}m_i(t) = \theta_i(t) -\epsilon \Delta^{(1)}_i(s) -\epsilon^2  \Delta^{(2)}_i(s) - \epsilon^3  \Delta^{(3)}_i(s) + {\cal O}(\epsilon^4)
\end{equation}
It is seen that to $\epsilon$ this is ``dynamic naive mean-field'', compare (\ref{eq:naive-mean-field}),
to $\epsilon^2$ this is ``dynamic TAP'', compare (\ref{eq:tap_dynamic}), and to
to $\epsilon^3$ typically there is a non-zero term absent in (\ref{eq:tap_dynamic}). Such a
higher-order difference between the exact mean-field theory for the asymmetric model and
the field-theoretical approach of~\cite{HertzRoudi2011,RoudiHertz2011} was also
pointed out in~\cite{Mezard2011} (page~4, in text below Eq.~7).

\section{Dynamic cavity method}
\label{sec:BP}
The cavity method for equilibrium systems was introduced in~\cite{MezardParisiVirasoro,MezardParisi2001} while the dynamic version
was studied but recently~\cite{KanoriaMontanari,NeriBolle,AurellMahmoudi2011}.  
In contrast to the equilibrium case, using only the cavity assumption does not in general
provide us with an efficient algorithm in the dynamic case, but further assumptions are necessary.
In this section we derive the dynamic cavity method for the kinetic Ising problem, taking
a more explicit route than in~\cite{NeriBolle} and~\cite{AurellMahmoudi2011}.

\subsection{Cavity and BP on spin histories}
We consider a number of spins evolving according to a dynamics
such as (\ref{eq:prob_dyn}), and we let $X_i$ denote the whole history of spin $i$,
$X_i=(\sigma_i(0),\sigma_i(1),\ldots,\sigma_T(0))$. 
The probability
in (\ref{eq:prob_dyn}) can then be alternatively be interpreted as a 
joint probability of spin histories, $P(X_1,X_2,\ldots,X_N)$, and this probability
can be represented by a graph where nodes $i$ and $j$ are
connected if either $J_{ij}$ or $J_{ji}$ (or both) are non-zero. The
corresponding product form is 
\begin{equation}
 P(X_1,X_2,\ldots,X_N)=\prod_ie^{\sum_s\theta_i(s)\sigma_i(s)} \prod_{ij}e^{\sum_s J_{ij}\sigma_i(s)\sigma_j(s+1)}
\prod_ie^{-\sum_s\log 2\cosh(\theta_i(s)+\sum_jJ_{ji}\sigma_j(s-1))}
\label{eq:product-form}
\end{equation}
which is already normalized.
Belief Propagation is expected to work well if this graph is locally tree-like
\textit{i.e.} if all loops are long, and can be ignored~\cite{MezardMontanari09}. 
In (\ref{eq:product-form}) this is never the case, even if the couplings are fully asymmetric, for the simple reason that if spins $i$ and $j$ 
drive spin $k$, then they are coupled both by the terms $\sigma_i(t)\sigma_k(t+1)$ and $\sigma_j(t) \sigma_k(t+1)$, and by the normalization 
$–\log 2\cosh h_k(t+1)$. However, these couplings are of a rather peculiar type.
To proceed we introduce four different marginal probabilities. The first $P_i$ is the marginal probability of spin history $X_i$. 
The second $P_{i+\partial i}$ is the marginal probability on the set of histories $\{X_i \bigcup X_{\partial i}\}$.  The third $P_{\partial i}$ is the marginal on 
the set of histories $X_{\partial i}$. The fourth and last is $P^{(i)}$, a cavity distribution on $X_{\partial i}$.
This we take as the marginal on $X_{\partial i}$ in a revised model where both the spin history $X_i$ as well as the 
normalization  $\log 2\cosh h_i(t)$ have been eliminated. All four probabilities depend on external field parameters which are not 
necessarily the same. In particular, we will express $P_{\partial i}$ with one set of external fields as $P^{(i)}$ with another set of external fields.
By definition $P_{i+\partial i}  =  W(X_i|\,X_{\partial i}) P_{\partial i}$. The peculiarity of the model is that the (normalized) conditional probability 
$W(X_i|\,X_{\partial i})$ is already explicitly included in (\ref{eq:product-form}). We can therefore compare
\begin{eqnarray}
P_{\partial i}(\vec{X}_{\partial i})&=&\frac{P_{i+\partial i}(X_i \bigcup X_{\partial i})}{ W(X_i| X_{\partial i})} \propto 
\prod_{j\in \partial i}e^{\sum_s\theta_j(s)\sigma_j(s)+\sum_{k\in \partial j}\sum_s J_{kj}\sigma_k(s)\sigma_j(s+1)-\sum_s\log 2\cosh(\theta_j(s)+\sum_kJ_{kj}\sigma_k(s-1))} \nonumber \\
&& \sum_{\vec{X}\setminus \{X_i \bigcup X_{\partial i}\}} 
\prod_{k\neq i,\partial i}\left( e^{\sum_s\theta_k(s)\sigma_k(s)} \prod_{l\in \partial k}e^{\sum_s J_{kl}\sigma_k(s)\sigma_l(s+1)} \prod_ke^{-\sum_s\log 2\cosh(\theta_k(s)+\sum_lJ_{lk}\sigma_l(s-1))}\right)
\label{eq:marginal1}
\end{eqnarray}
to
\begin{eqnarray}
P^{(i)}(X_{\partial i})&\propto& \prod_{j\in \partial i}e^{\sum_s\theta_j(s)\sigma_j(s)+\sum_{k\in \partial j, k\neq i}\sum_s J_{kj}\sigma_k(s)\sigma_j(s+1)-\sum_s\log 2\cosh(\theta_j(s)+\sum_{k\neq i}J_{kj}\sigma_k(s-1))} \nonumber \\
&& \sum_{\vec{X}\setminus \{X_i \bigcup X_{\partial i}\}} 
\prod_{k\neq i,\partial i}\left(e^{\sum_s\theta_k(s)\sigma_k(s)} \prod_{l\in \partial k}e^{\sum_s J_{kl}\sigma_k(s)\sigma_l(s+1)} \prod_ke^{-\sum_s\log 2\cosh(\theta_k(s)+\sum_lJ_{lk}\sigma_l(s-1))}\right)
\label{eq:marginal2}
\end{eqnarray}
This comparison shows that $P_{\partial i}$ with external fields $\theta_j(t)$ is the same as $P^{(i)}$ with {\it modified} external fields 
$\theta_j(t) + J_{ij} \sigma_i(t-1)$.
Since $P_i(X_i)=\sum_{X_{\partial i}}  W(X_i|\,X_{\partial i}) P_{\partial i}(X_{\partial i})$
we can therefore write the marginal probability $P_{i}$ as

\begin{eqnarray}
P_i(X_i\,|\,\theta_i(0),...,\theta_i(t),\cdot) = &\displaystyle\sum_{\sigma_{\partial i}(0)...\sigma_{\partial i}(t-1)}\,
P^{(i)}(X_{\partial i}(0)\,| {\theta}^{\,(i)}_{\partial i}(0),\ldots,{\theta}^{\,(i)}_{\partial i}(t),\cdot)&\nonumber\\
&\displaystyle\prod_{s=1}^{t}\,W_i(\sigma_i(s)\,|\,h_i(s))\,\, p_i(\sigma_i(0))&
\label{eq:mar_cav}
\end{eqnarray}
where $\cdot$ indicates all the parameters (external fields and interactions) which are
the same on the two sides of the equation, and 
\begin{equation}
\theta^{(i)}_j(s)= \theta_j(s) + J_{ij} \sigma_i(s-1) \qquad s=0,\ldots,t
\end{equation}
is the set of external fields that are modified.

The next step is to make the Belief Propagation assumption that the
spin histories are taken independent in the cavity graph:
\begin{equation}
P^{(i)}(X_{\partial i}\,| \theta^{\,(i)}_{\partial i}(0),\ldots,\theta^{\,(i)}_{\partial i}(t),\cdot)
 =  \prod_{j\in \partial i}  \,\, 
\mu_{j\to i}(X_j|\theta^{(i)}_{j}(0),\ldots,\theta^{(i)}_{j}(t),\cdot) 
\label{eq:fact}
\end{equation} 
We here used used $\mu_{i\to j}$ to represent marginal probabilities of neighboring spins,
as standard in the BP literature. 

We now consider the subgraph $T^{(i)}_j$ connected to one spin $j$ in the cavity of $i$.
In analogy to above we want to compare the marginal on the set of neighbours to spin $j$
in $T^{(i)}_j$, to the cavity distribution on the same set of variables. As above
the first with one set of external fields is the same as the second with another set 
of external fields, and we therefore find the following
recursion equations (``BP update equations''):
\begin{eqnarray}
\mu_{j\to i}(X_j|\theta^{(i)}_j(0),...,\theta^{(i)}_j(t),\cdot_{T^{(i)}_j}) =\sum_{X_{\partial j\setminus i}}  &\displaystyle\prod_{k\in\partial j\setminus i}&\mu_{k\to j}\left(X_k|\theta^{(i),(j)}_k(0),...,\theta^{(i),(j)}_k(t-1),\cdot_{T^{(i)}_j}\right) \nonumber\\
&&\displaystyle\prod_{s=1}^{t} w_j(\sigma_j(s)\,|\,h_j^{(i)}(s))\,\,\,\,\mu_{j\to i}(\sigma_j(0))
\label{eq:bp_mess_syn}
\end{eqnarray}
where
$\cdot_{T^{(i)}_j}$ indicates all the parameters (external fields and interactions) which are
the same on the two sides of the equation, and 
\begin{equation}
\theta^{(i),(j)}_k(s) = \theta^{(i)}_k(s)+J_{jk}\sigma_j(s-1) \qquad s=0,\ldots,t 
\label{eq:external-fields}
\end{equation}
is the set of external fields that are modified. 
Since in fact $\theta^{(i)}_k(s)=\theta_k(s)$ (spin $k$ is not directly connected to $i$)
we note that in  (\ref{eq:external-fields}) the upper index $(i)$ can be dropped on both sides.
The effective field on spin $j$ in $T^{(i)}_j$ is
\begin{equation}
h_j^{(i)}(s) = \sum_{k\in\partial j\setminus i} J_{kj}\,\sigma_k(s-1) + \theta_j(s)
\label{eq:cavity-field-def}
\end{equation}
and $w_j(\sigma_j\,|\,h_j^{(i)}(s))$ is the transition probability for the single spin $j$ 
in the model on $T^{(i)}_j$.

%
%
%

%
The marginal probability over the history of one spin (``BP output equation'') follows from
(\ref{eq:mar_cav}) and (\ref{eq:fact}) and is 
\begin{eqnarray}
P_i(X_i\,|\,\theta_i(0),...,\theta_i(t),\cdot) = &\displaystyle\sum_{X_{\partial i}}\,
\displaystyle\prod_{k\in\partial i}\mu_{k\to i}(\sigma_k(0),...,\sigma_k(t-1)|\theta^{(i)}_k(0),...,\theta^{(i)}_k(t-1),\cdot)&\nonumber\\
&\displaystyle\prod_{s=1}^{t}\,W_i(\sigma_i(s)\,|\,h_i(s))\,\, p_i(\sigma_i(0))&
\label{eq:bp_output}
\end{eqnarray}
Equations (\ref{eq:bp_mess_syn}) and (\ref{eq:bp_output}) are the dynamic cavity 
equations for our system. Both are large sets of equations connecting marginal distributions 
and cavity distributions between two probabilistic models with different parameters.
In general these equations are (as far as we know) 
only of conceptual value since on top of connecting different models, the right hand side also involves
on the order of $2^{T|\partial i|}$ operations. In BP such an operation would 
have to be iterated (for all variables)
a number of times to reach convergence: as $T$ grows large this becomes unfeasible for the same reason
that ordinary BP does not work well if the state space of each variable is large. 

We can define (and will later use) marginalizations of the messages down to one time
(it is no restriction to take this time as the last time):
\begin{equation}
\mu_{j\to i}^t(\sigma_j(t)|\theta_j^{(i)}(t)) = \sum_{\sigma_j(0)\ldots \sigma_j(t-1)}\mu_{j\to i}(\sigma_j(0),...,\sigma_j(t)|\theta^{(i)}_j(0),...,\theta^{(i)}_j(t)) 
\label{eq:single_message}
\end{equation}
but in general these quantities do not obey closed equations among themselves.
An important exception are fully 
asymmetric networks, since there at most one of $J_{ij}$ and $J_{ji}$ 
is non-zero. We note that in (\ref{eq:bp_mess_syn}) and (\ref{eq:bp_output}) 
the probability distribution of spin $i$ depends on the
neighbors $\partial i$ through the effective fields 
$h_j^{(i)}(s)$ and $h_i(s)$, but the messages sent from 
the neighbors to $i$ also depend parametrically on the history of $i$ through the
modified external fields $\theta^{(i)}_k$. This back-action is absent for 
the fully asymmetric case where $\theta^{(i)}_k = \theta_k$ independent of spin $i$.

 \subsection{The projected dynamic BP}
As discussed the marginalization of dynamic BP over one time 
is not in general a Markov process.
However, the long time behavior of dynamics (stationary state) is often demanded in many cases. 
In this section we explain an approximation scheme for computing marginal probabilities 
for one spin over one time in stationary state, a procedure 
called \textit{one-time approximation} in~\cite{NeriBolle} and 
\textit{time factorization} in~\cite{AurellMahmoudi2011,AurellMahmoudi-2011b}.

We start with the dynamic BP for the time histories of messages, {\it i.e.}~Eq(\ref{eq:bp_mess_syn}),
where on the right hand side the time trajectory of messages sent from neighboring spins carry the 
information from the whole time history of those spins. 
We note that the full dynamics Eq.~(\ref{eq:prob_dyn}) is in fact Markov.
This, and the need to introduce some approximation, motivates 
\textit{the time-factorization ansatz} which we write for the terms in the right-hand side of~Eq(\ref{eq:bp_mess_syn}):
\begin{equation}
\mu_{k\to j}\left(\sigma_k(0),\ldots,\sigma_k(t)\,| \cdot_{T^{(j)}_k}\right)  = \prod_{s=0}^t\,\, \mu^s_{k\to j}\left(\sigma_k(s)\,| \cdot_{T^{(j)}_k}\right)
\label{eq:time_factorization}
\end{equation}
where $\cdot_{T^{(j)}_k}$ indicates the parameters of the model,
the same on both sides.
Obviously, when inserted into the right-hand side of~(\ref{eq:bp_mess_syn})
such a factorization is not preserved on the left hand side.
Since we deal 
with binary variables we can introduce
{\it time-factorized cavity biases} $u_{k\to j}^{{T^{(j)}_k}}(s)$, again written
for the right-hand side of (\ref{eq:bp_mess_syn}) which are defined by
\begin{equation}
\mu^s_{k\to j}\left(\sigma_k(s)\,|\,\cdot_{T^{(j)}_k} \right) = 
\frac{e^{\beta \left[\,u_{k\to j}^{T^{(j)}_k}(s)\,\sigma_k(s)\right]}}
{2\cosh \left[\beta \left(u_{k\to j}^{T^{(j)}_k}(s)\right )\right]}
\label{eq:time-factorized-cavity-fields}
\end{equation}
A crucial observation is now that when the time-factorization ansatz has been made 
the cavity biases at different external
fields are simply related. We will need
\begin{equation}
u_{k\to j}^{T^{(j)}_k}(s)=u_{k\to j}^{T^{(i)}_j}(s) + J_{jk}\sigma_j(s-1) \qquad s=0,\ldots,t
\label{eq:time-factorized-cavity-fields-different-external-fields}
\end{equation}
which follows from the relation (\ref{eq:external-fields}).  
Inserting (\ref{eq:time-factorized-cavity-fields}) 
and (\ref{eq:time-factorized-cavity-fields-different-external-fields})
into~(\ref{eq:bp_mess_syn}) gives
\begin{eqnarray}
\mu_{j\to i}(\sigma_j(0),...,\sigma_j(t)|\, \cdot_{T^{(i)}_j}) =
\sum_{\sigma_{\partial j\setminus i}(0),...,\sigma_{\partial j\setminus i}(t-1)} &  \displaystyle\prod_{s=0}^t\displaystyle\prod_{k\in\partial j\setminus i} \frac{e^{\beta \sigma_k(s)\left(\,u_{k\to j}^{T^{(i)}_j}(s)\,+J_{jk}\sigma_j(s-1)\right)}}
               {2\cosh \left[\beta\left(\,u_{k\to j}^{T^{(i)}_j}(s)\,+J_{jk}\sigma_j(s-1)\right) \right]}&\nonumber\\
&\displaystyle\prod_{s=1}^{t} w_j(\sigma_j(s)\,|\,h_j^{(i)}(s))\,\,\,\,\mu_{j\to i}(\sigma_j(0))&
\label{eq:bp_mess_time_fact}
\end{eqnarray}
This equation can be marginalized explicitly over the last time to give
\begin{equation}
\mu^t_{j\to i}(\sigma_j(t)|\, \cdot_{T^{(i)}_j}) = \sum_{\sigma_{j}(t-2),\sigma_{\partial j\setminus i}(t-1)} \displaystyle\prod_{k\in\partial j\setminus i}\frac{e^{\beta \sigma_k(s)\left(u_{k\to j}^{T^{(i)}_j}(t-1)+J_{jk}\sigma_j(t-2)\right)}}
               {2\cosh \left[\beta\left(u_{k\to j}^{T^{(i)}_j}(t-1)+J_{jk}\sigma_j(t-2)\right) \right]}\,w_j(\sigma_j(t)| h_j^{(i)}(t))\,\mu^{t-2}_{j\to i}(\sigma_j(t-2)|\, \cdot_{T^{(i)}_j})
\label{eq:stationary_mess}
\end{equation} 
The \textit{projected dynamic cavity} is then to use (\ref{eq:stationary_mess}) to
compute the terms in a time-factorization of the left-hand side of~Eq(\ref{eq:bp_mess_syn}).
Except for fully asymmetric models (with parallel updates), this approach is not 
appropriate for transients~\cite{AurellMahmoudi-2011b}.
However, when the external fields $\theta_i$ are
constant in time and when a stationary state has been reached,
it may be acceptable to also take the messages independent
in time.
For one and the same set of parameter values the fixed-point equations for
the time-independent time-factorized cavity biases are then
\begin{eqnarray}
u_{j\to i}^{*} = \frac{1}{2\beta}\sum_{\sigma_j}\,\sigma_j\log\left[\sum_{\sigma_{\partial j\setminus i},\sigma_j'}\frac{e^{\beta \sum_{k\in\partial j\setminus i}\sigma_k\left(u_{k\to j}^{*}+J_{jk}\sigma_j'\right)}}
               {2\cosh \left[\beta\left(u_{k\to j}^{*}+J_{jk}\sigma_j'\right) \right]}\right.\nonumber\\
\left.\frac{e^{\beta\,h_j^{(i)}\sigma_j}}{2\cosh(\beta\,h_j^{(i)})}\,\,\frac{e^{\beta \sigma_j' 
u_{j\to i}^{*}}}
               {2\cosh \left(\beta u_{j\to i}^{*}\right)}\right]
\qquad h_j^{(i)} = \sum_{k\in\partial j\setminus i} J_{kj}\,\sigma_k + \theta_j
\label{eq:cavity-bias-eq}
\end{eqnarray}
Eq (\ref{eq:cavity-bias-eq}) is as ordinary BP solved by iteration,
where the right-hand side is computed from
$u_{k\to j}^{(t-1)}$ at iteration time $t-1$,
giving the left hand side $u_{j\to i}^{(t)}$ at iteration time $t$.
The spin $\sigma_j$ summed over is then conceptually at time $t$,
the spins $\sigma_k$ at time $t-1$ and the last spin $\sigma_j'$ at time $t-2$,
all these in the iteration time.

Using the iteration time as a proxy for real time we
note that in a transient we can
compute the time evolution of magnetization which would follow from 
(\ref{eq:time-factorized-cavity-fields}), (\ref{eq:cavity-field-def})
\begin{eqnarray}
m_i(t) = \sum_{\sigma_{\partial i\setminus j}(t-1),\sigma_i(t-2)}\,\frac{e^{\beta \sum_{k\in \partial i\setminus j}\left[u_{k\to i}(t-1)+J_{ik}\sigma_i(t-2)\right]\sigma_k(t-1)}}{\prod_{{k\in \partial i\setminus j}}2\cosh[\beta \left(u_{k\to i}(t-1)+J_{ik}\sigma_i(t-2)\right)]}\nonumber\\
\tanh\left[\beta(\sum_{j\in\partial i}J_{ji}\,\sigma_j(t-1)+\theta_i)\right]\,\frac{e^{\beta u_{i\to j}(t-2)\sigma_i(t-2)}}{2\cosh(\beta u_{i\to j}(t-2))}\
\label{eq:cavity-magnetizations}
\end{eqnarray}
This is not expected to be accurate unless we are already in a stationary state.
We use it below in Section~\ref{sec:results} as a proxy to monitor if the system is 
in a stationary state.

\section{Results}
\label{sec:results}
In this section we investigate the performance of dynamic cavity method in computing stationary states of diluted spin glass 
in parallel update, and compare to MCMC (Glauber dynamics) and to dynamic mean-field and dynamic TAP as
defined in Section~\ref{sec:infinite_range}.
The convergence of projected dynamic cavity (dynamic cavity in time-factorized approximation)
is monitored by comparing magnetization computed from 
(\ref{eq:cavity-magnetizations}) at successive times for different parameter values of the model,
and these predictions are then compared to dynamic mean-field and dynamic TAP and MCMC.
\subsection{Convergence of dynamic BP}
\begin{figure}[htb]
\vspace{0.2cm}
\includegraphics[width=0.43\columnwidth]{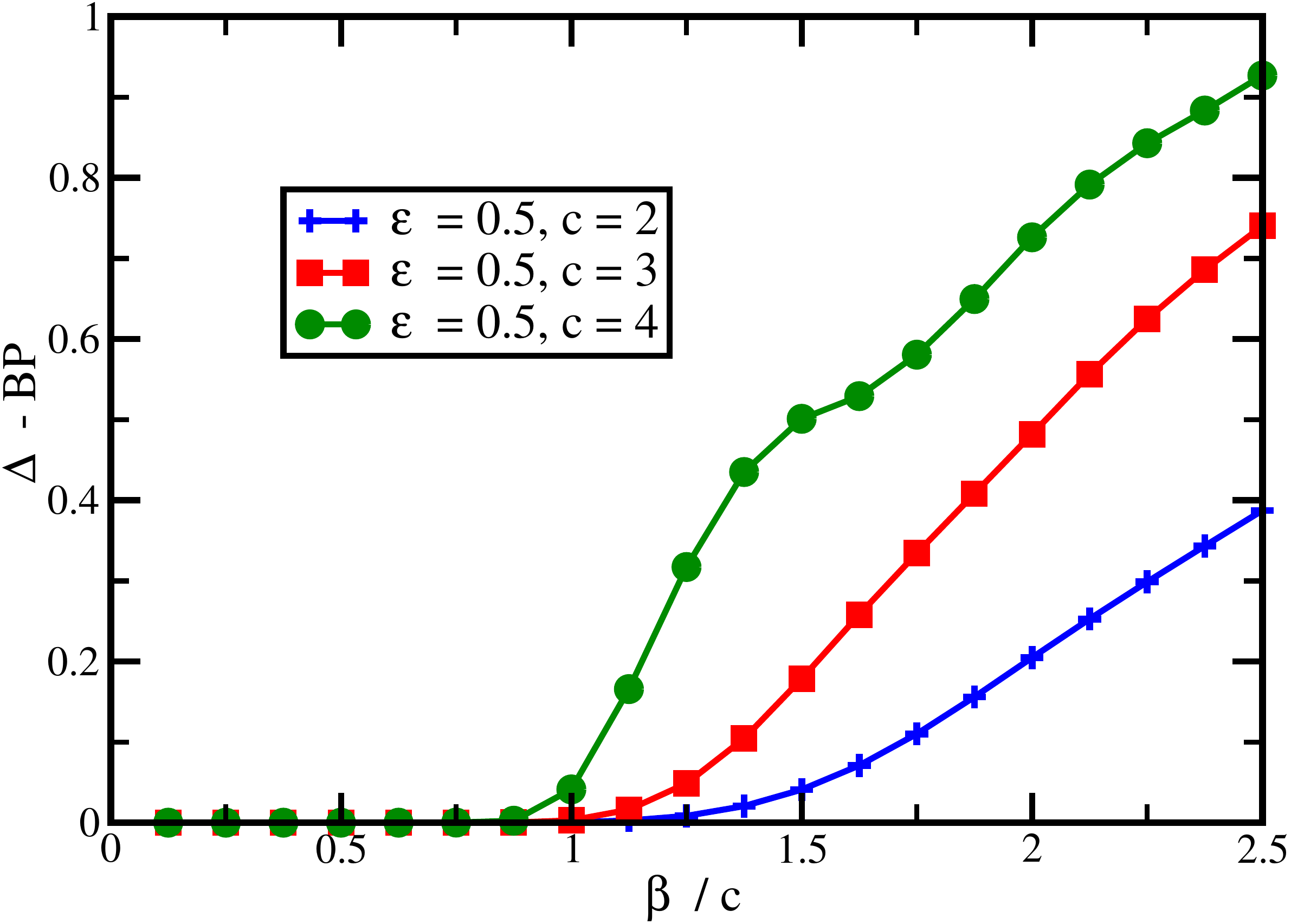}
\hspace{10mm}
\includegraphics[width=0.43\columnwidth]{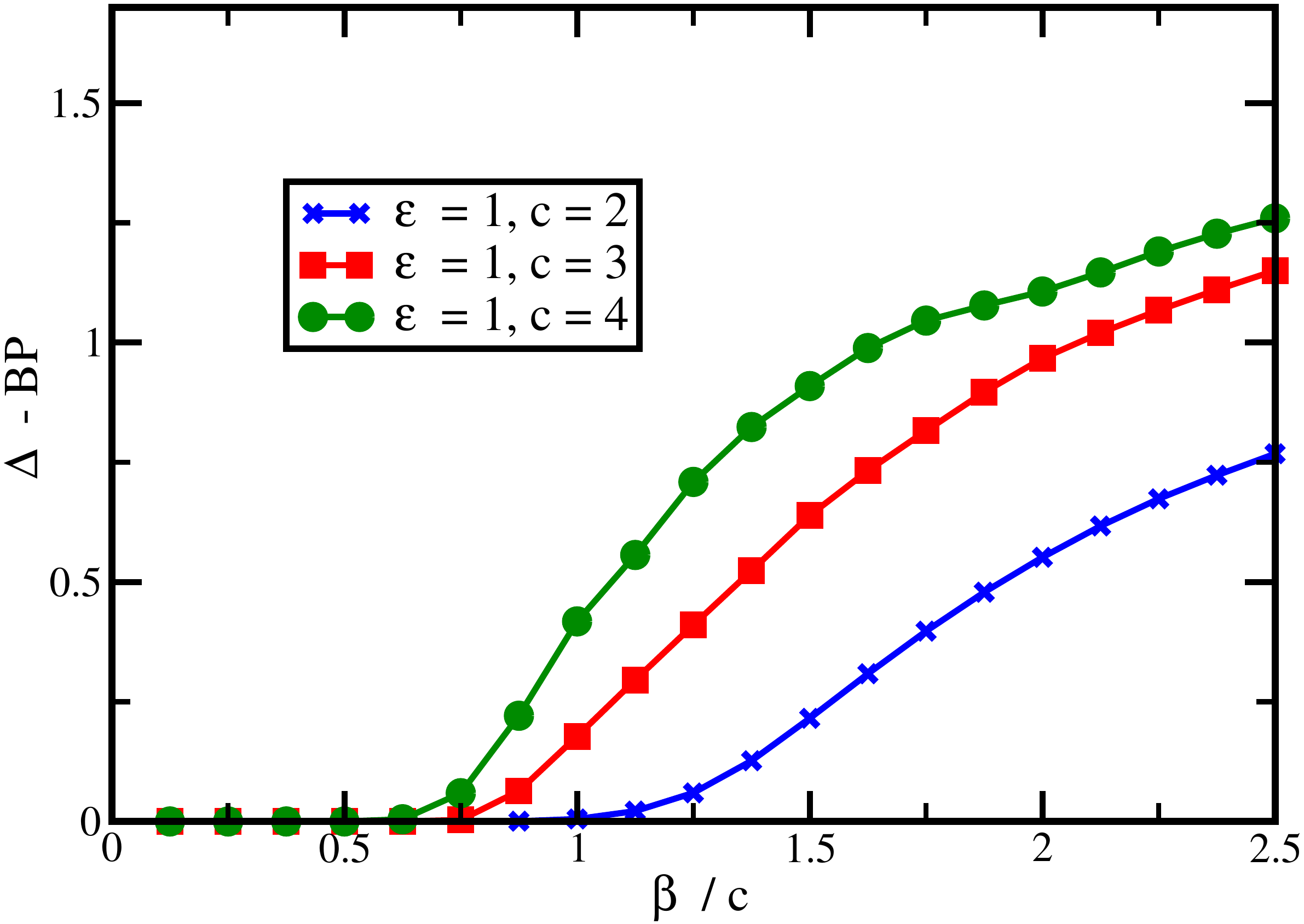}
\caption{(Color online) Squared deviation of  spin averages between successive update $\Delta (t) = 1/N\sum_{i=1}^N\, (m_i(t)-m_i(t-1))^2$ at stationary limit
for different values of average connectivity $c$. Mean magnetizations are calculated by projected dynamic cavity method {\it i.e} Eq.~\ref{eq:cavity-magnetizations}.
Left panel: partially asymmetric networks with $\epsilon = 0.5$. Right panel : symmetric network $\epsilon = 1$. The results are averaged over BP initial conditions (10 experiences).
System size is $1000$ and external fields are set to zero. }
\label{fig:converg1}
\end{figure}

\begin{figure}[htb]
\vspace{0.6cm}
\includegraphics[width=0.32\columnwidth]{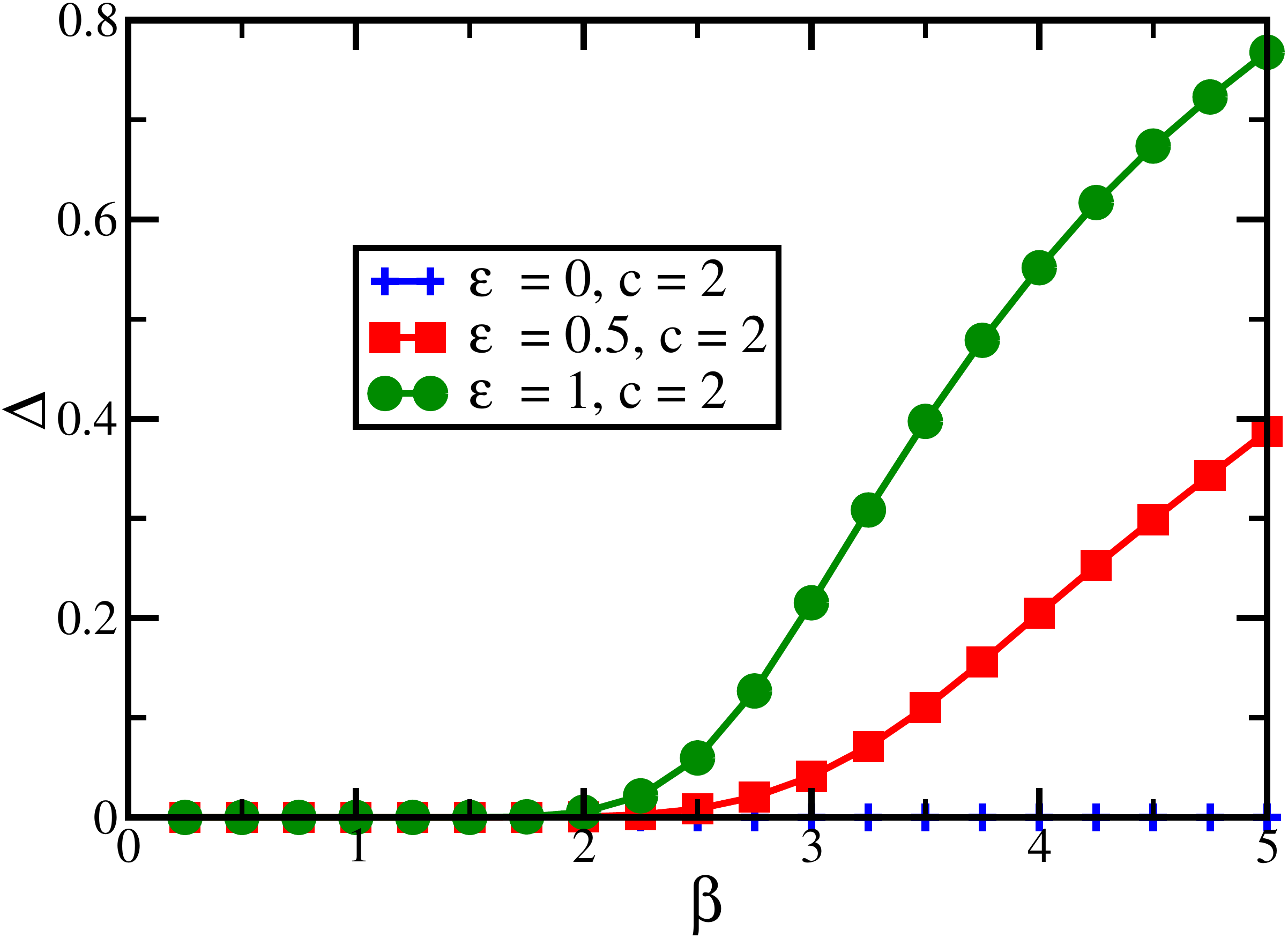}
\hspace{0.5mm}
\includegraphics[width=0.32\columnwidth]{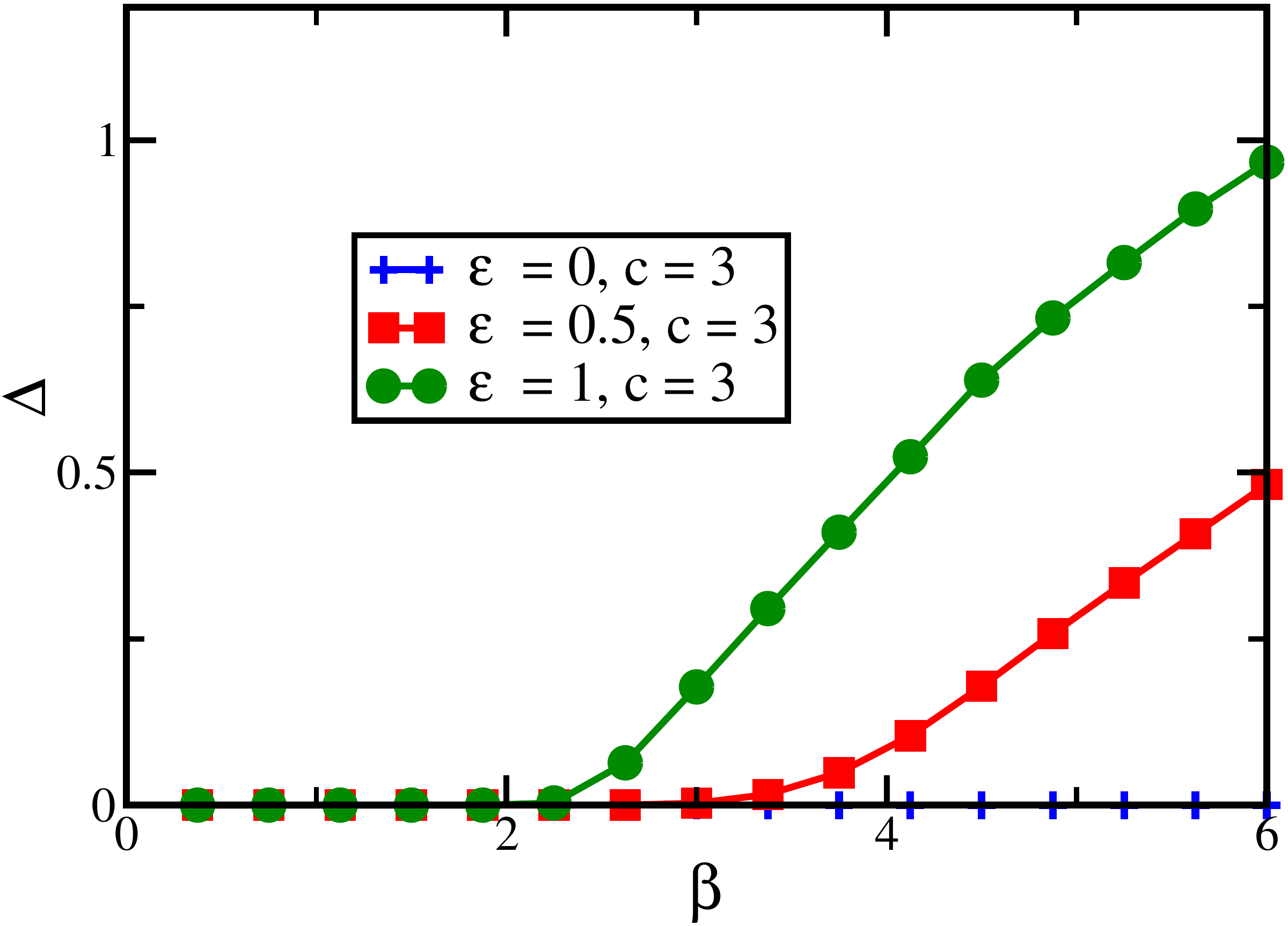}
\hspace{0.5mm}
\includegraphics[width=0.32\columnwidth]{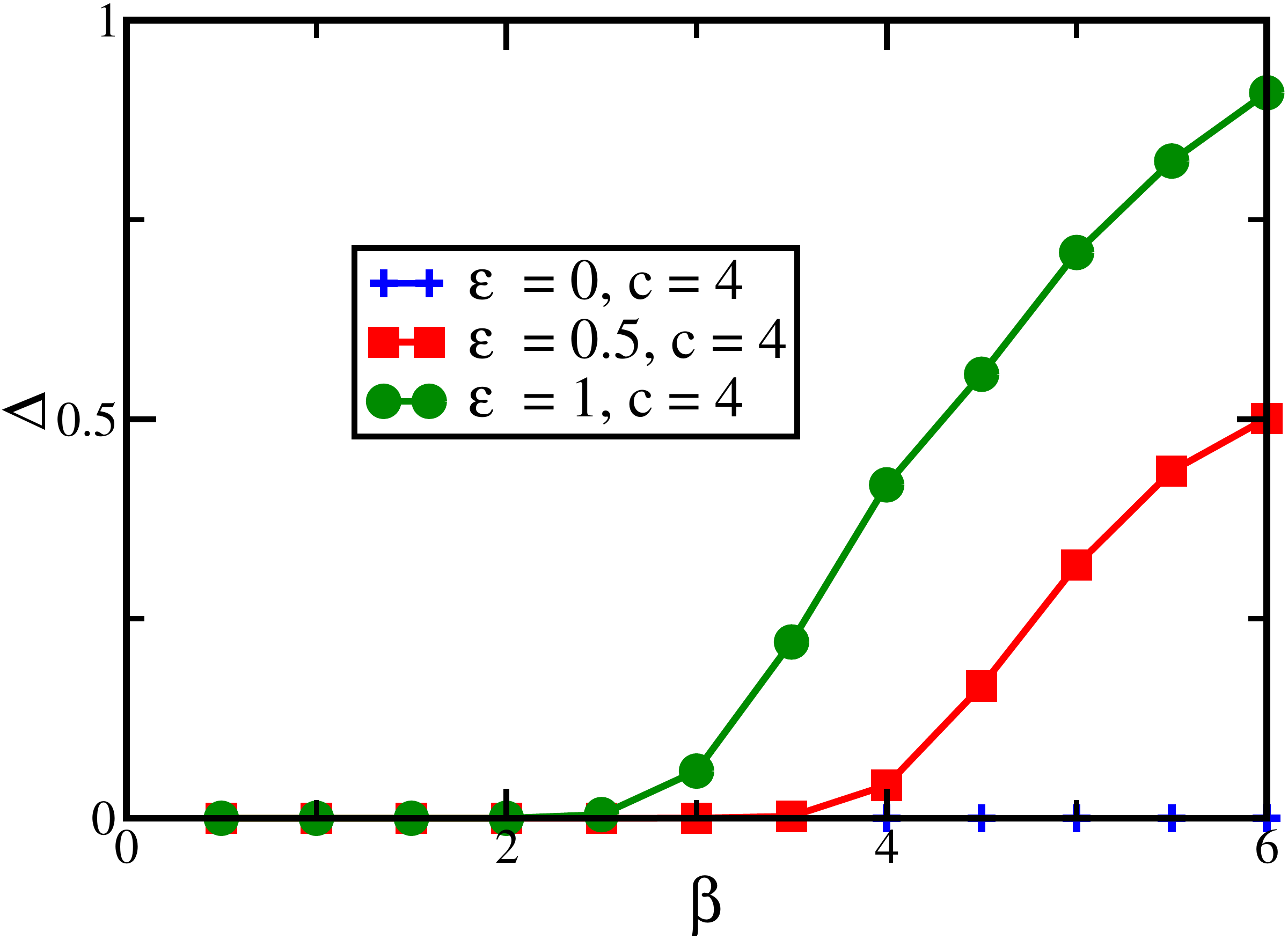}
\caption{ (Color online) Effect of asymmetry $(\epsilon = 0, 0.5, 1)$ in squared deviation of  spin averages between successive update $\Delta (t) = 1/N\sum_{i=1}^N\, (m_i(t)-m_i(t-1))^2$, obtained by projected dynamic BP Eq.~\ref{eq:cavity-magnetizations}, at stationary limit
for average connectivity $c=2$ (left panel), $c=3$ (middle panel), $c=4$ (right panel). The results are averaged over BP initial conditions (10 experiences).
System size is $1000$ and external fields are set to zero.
 }
\label{fig:converg2}
\end{figure}

In order to detect where dynamic BP reaches a stationary state we compare single magnetization in two successive time step
as
\begin{equation}
\Delta(t) = 1/N \sum_{i=1}^N\, \left(m_i(t)-m_i(t-1)\right)^2
\end{equation}
Whenever this deviation vanishes dynamic BP must have converged to a stationary state. 
Fig.~\ref{fig:converg1}
shows the results for various connectivity parameters in symmetric and partially symmetric networks.
In high temperature we observe convergence towards a fixed point whereas in low temperature BP does not reach a stationary state.
Roughly speaking, dynamic BP stops converging at a value $\beta_{cr}(c)$ which depends on average connectivity.
In Fig.~\ref{fig:converg2} the convergence of dynamic BP is plotted to show the effect of asymmetry.
In this case it is simply so that for very asymmetric graphs BP converges in a very wide region, presumably
for arbitrarily large values of $\beta$ if the network grows large enough, and, in general, the
more asymmetric the network, the better the convergence. 

\subsection{Performance of dynamic BP}
\begin{figure}[htb]
\vspace{0.2cm}
\includegraphics[width=0.43\columnwidth]{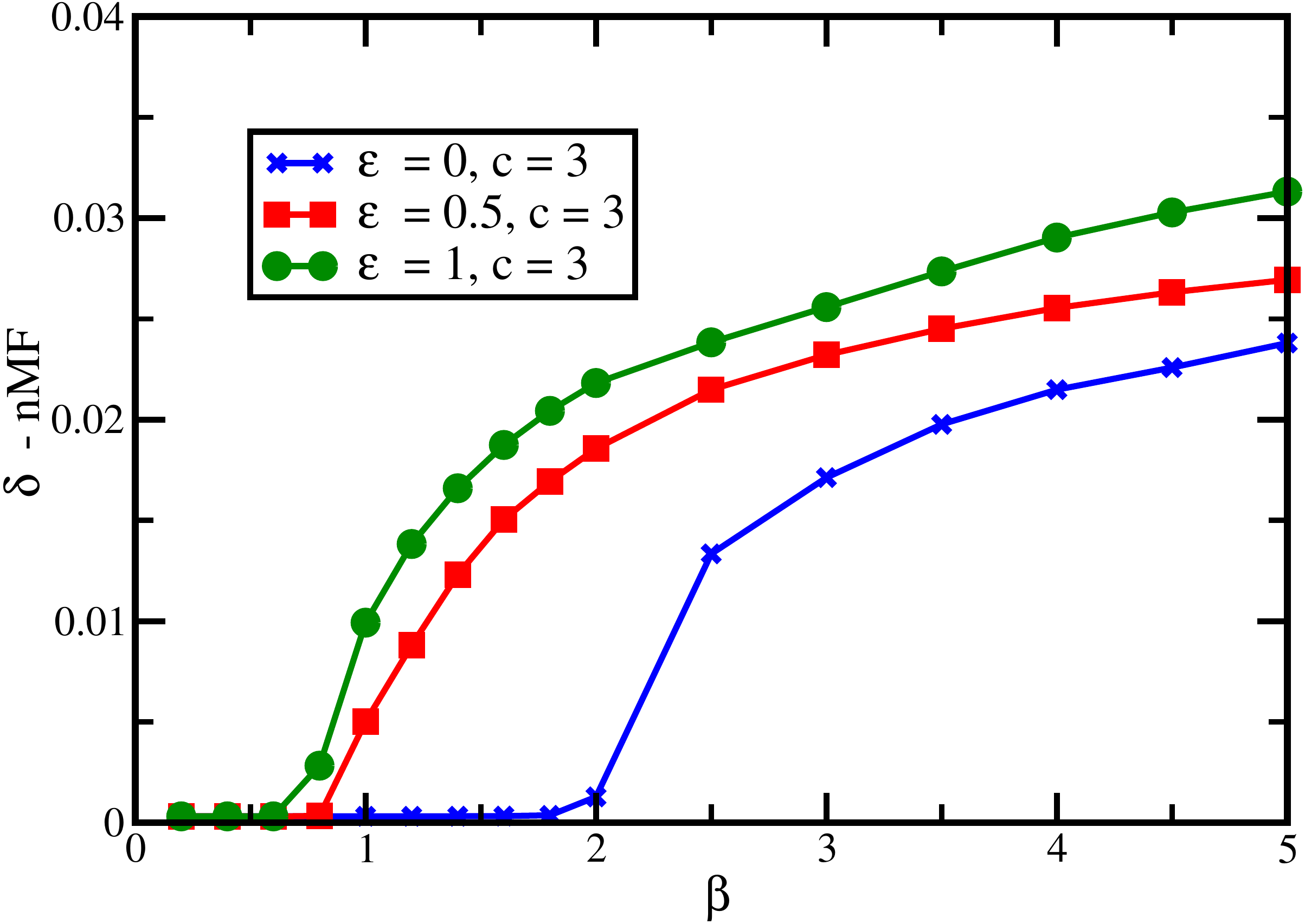}
\hspace{10mm}
\includegraphics[width=0.43\columnwidth]{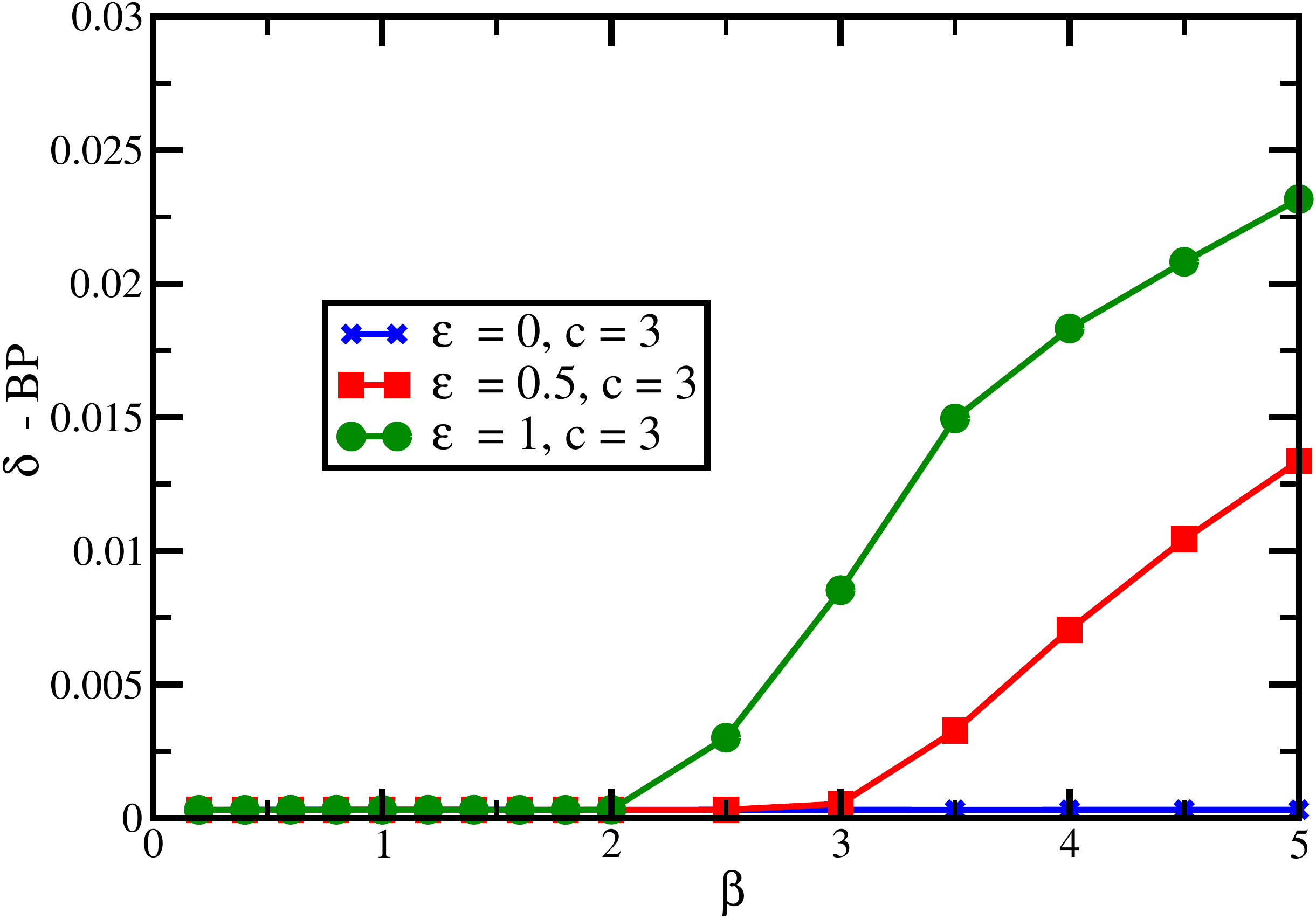}
\caption{(Color online) Mean square error $\delta (t) = 1/N \sum_{i=1}^N (m_i^{\rm predicted}(t)-m_i^{\rm empirical}(t))^2$ of the two approximation methods (dynamic mean field and dynamic cavity with respect to the empirical data (Glauber dynamics). 
Left panel: dynamic mean field (Eq.~\ref{eq:naive-mean-field}) for networks with different asymmetric parameter $(\epsilon =0, 0.5, 1)$ and fixed average connectivity $c=3$. Right panel : the corresponding results obtained by the projected dynamic cavity method Eq.~\ref{eq:cavity-magnetizations}) For small $\beta$, {\it i.e.} high temperature they
are in agreement with numerical simulations. In low $\beta$ however, dynamic BP outperforms dynamic mean field.}
\label{fig:nMF_BP}
\end{figure}

Fig.~\ref{fig:nMF_BP} shows a comparison between dynamic cavity method and dynamic mean field for total magnetization
in spin glass systems with different asymmetric parameter. The results are obtained in present of small external fields $\theta=0.001$. Dynamic cavity method 
shows a strong agreement with numerical simulations of type Glauber dynamics when it converges to a stationary state. The dynamic mean field method however starts to deviate from 
numerical simulations already in small $\beta$ indicating that it is less accurate compared to the dynamic cavity method. 

\begin{figure}[htb]
\vspace{0.2cm}
\includegraphics[width=0.45\columnwidth]{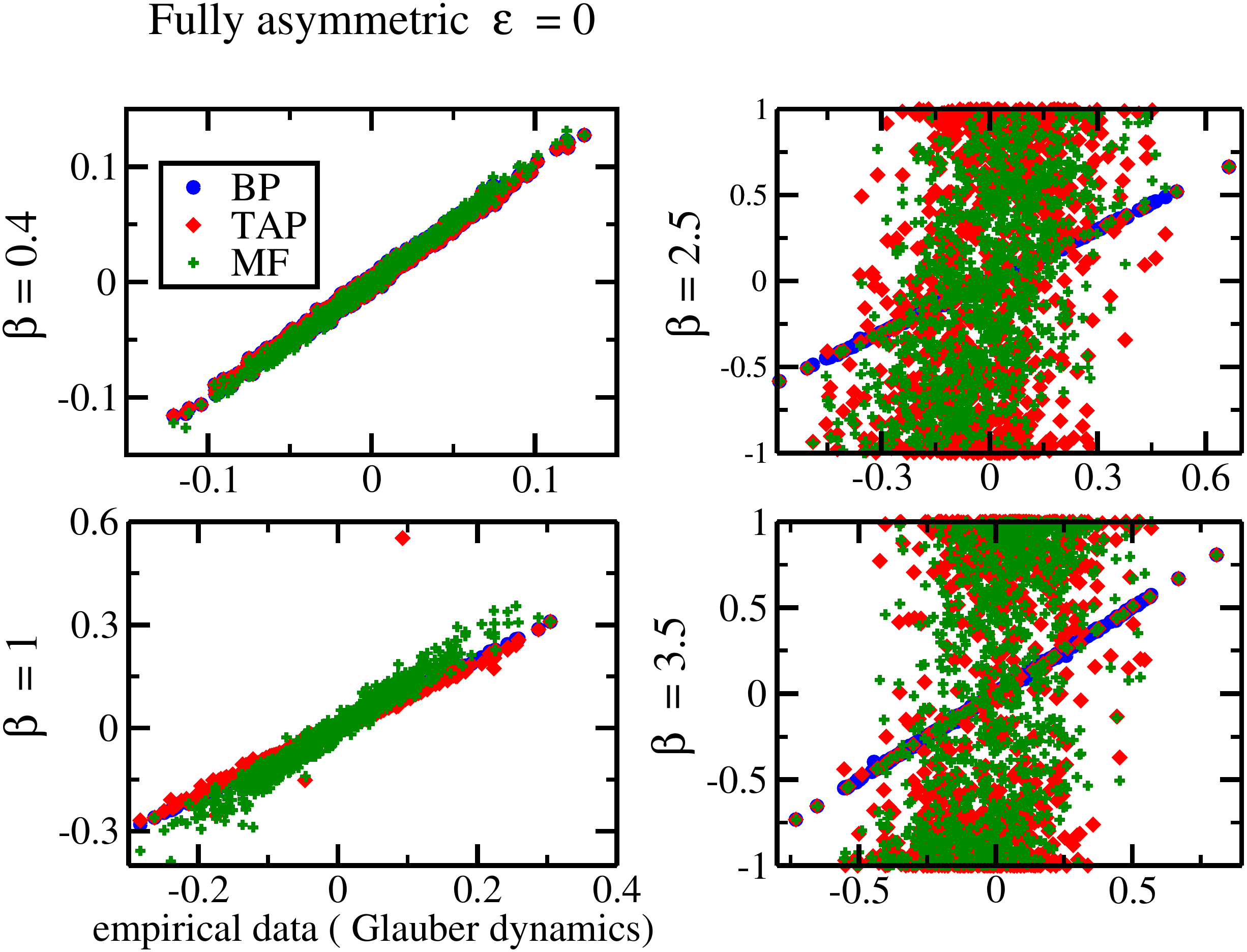}
\hspace{10mm}
\includegraphics[width=0.45\columnwidth]{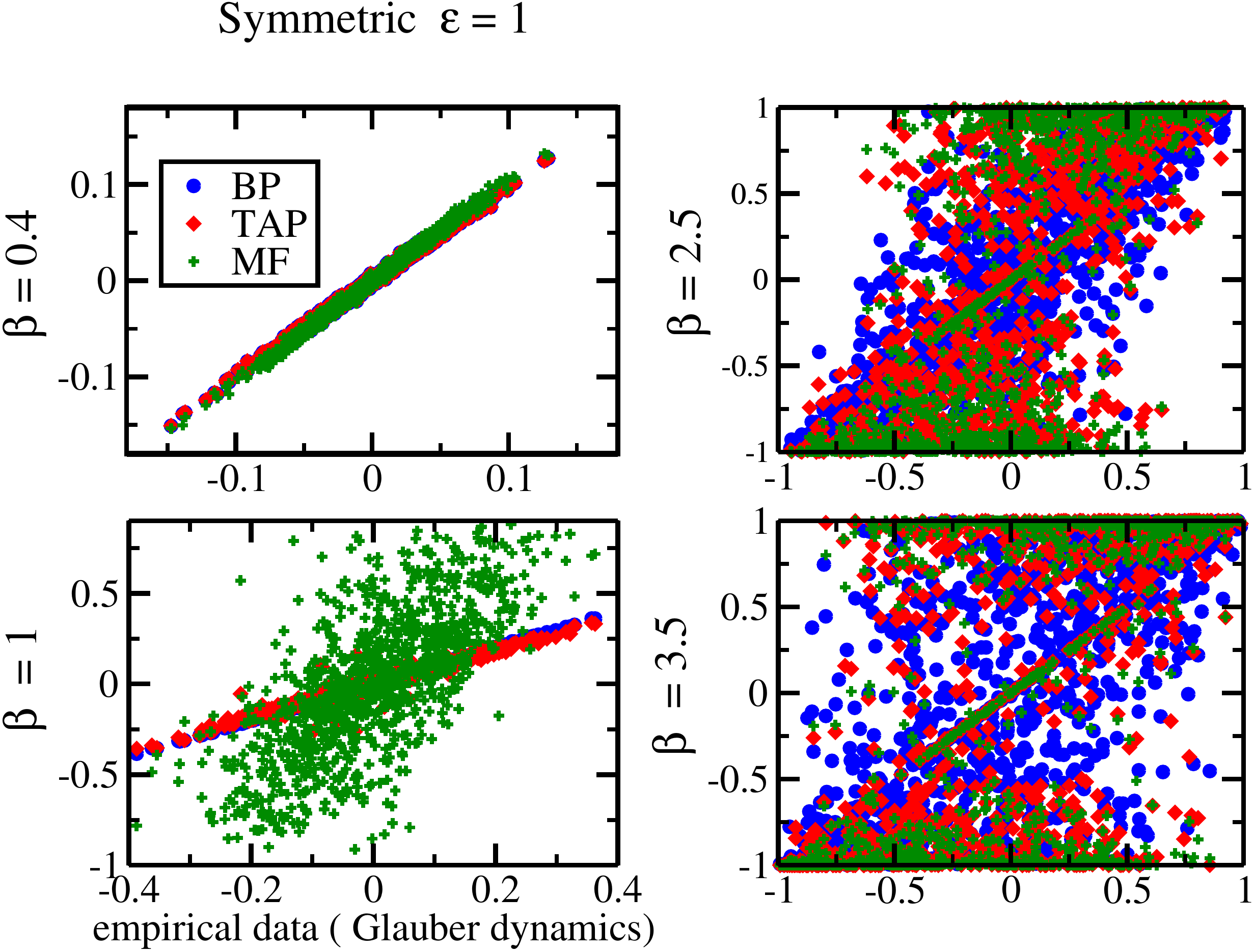}
\caption{(Color online)
Scatter plot of local magnetizations for dilute asymmetric networks for four different temperature $\beta=0.4, 1, 2.5, 3.5$  and a fixed average connectivity $c=3$.
Local magnetizations are obtained by dynamic mean field Eq.~\ref{eq:naive-mean-field} (green), dynamic TAP Eq.~\ref{eq:tap_dynamic} (red), and projected dynamic BP Eq.~\ref{eq:cavity-magnetizations} (blue).
Left panels show the scatter plots in fully asymmetric networks ($\epsilon=0$) where the projected dynamic BP provides exact results and right panels are scatter plots in fully symmetric networks $(\epsilon=1)$. In high temperature, all three methods agree with 
numerical simulations. In low temperature BP starts to outperform naive mean field and TAP.}  
\label{fig:scatter1}
\end{figure}

In order to observe the comparison in more detail, we show also the scatter plot of spin-by-spin magnetization in Fig.\ref{fig:scatter1}. 
Dynamic cavity method predicts perfectly local magnetizations for fully asymmetric networks and agrees quite well with numerical simulations in high temperature for fully symmetric network whereas 
naive mean field and TAP start to deviate already at moderate temperatures. 

\section{Conclusion}
\label{sec:conclusion}
Message-passing methods have become an important topic on the border-line
between equilibrium statistical physics and information theory. In the present paper we
have studied an extension of message-passing to non-equilibrium Ising spin systems.
In contrast to the equilibrium case, the cavity method is not immediately useful 
to describe the dynamics, even if the topology is suitable, because the messages
depend on whole spin time histories. The time-factorization assumption,
as discussed here and in~\cite{NeriBolle,AurellMahmoudi2011,AurellMahmoudi-2011b}, 
(or some other simplifying assumption) is necessary to reduce the complexity,
but when so doing one is generally restricted to stationary states.

We have studied dynamic cavity in the time-factorized assumption for stationary
states and outlined its convergence region in parameter strength ($\beta$),
connectivity ($c$) and asymmetry ($\epsilon$). 
By analogy with generally known facts
about BP it can be argued that when dynamic
cavity converges it should typically be a good approximation; the region of convergence
is therefore a useful proxy for the accuracy. 
Expanding on first results presented
in~\cite{AurellMahmoudi2011} we show that the convergence region in $\beta$
\textit{increases} with the connectivity.
We also find that the convergence region \textit{increases} with asymmetry for several
values of connectivity, and that it converges for any interaction strength for
fully asymmetric networks (as expected). For networks of moderate size we
have directly compared dynamic cavity and dynamic mean-field to direct simulation.
For several values of asymmetry and connectivity we find that their convergence 
regions are very similar, if not identical, but when both methods converge, then
dynamic cavity is considerably more accurate, except in the low $\beta$ limit
where their performance is about the same.
We have hence showed that dynamic cavity can be useful new approximation to the
dynamics of non-equilibrium spin systems -- and any system which can be fruitfully
modeled by such methods.

On the analytical side we have discussed the special status of fully asymmetric
models, for which the cavity approach is in some sense exact. We have also
re-derived the ``dynamic TAP'' equation of 
Hertz and Roudi~\cite{HertzRoudi2011,RoudiHertz2011} using a straight-forward approach
borrowed from Kappen and Spanjers' treatment of the stationary state~\cite{KappenSpanjers2000}
clarifying that this approach is based on minimizing the distance, in the sense
of Information Geometry, to the sub-family of independent (but time-changing) models.
Whether such a perturbative argument can be extended to small deviations
from \textit{e.g.} fully asymmetric models remains to be seen. 

\section*{Acknowledgment}
We thank Marc M\'ezard for important remarks, and 
Mikko Alava, Yoshiyuki Kabashima, Pekka Orponen and Toshiyuki Tanaka for discussions.
The work was supported by the
Academy of Finland as part of its Finland Distinguished
Professor program, project 129024/Aurell.

\appendix
\section{The Information Geometry calculation to second order}
The following calculations are completely parallel to
those in Appendix~1 of~\cite{KappenSpanjers2000} and start from
\begin{eqnarray}
\frac{\partial m_i(t)}{\partial \theta_j(s)}|_{\hbox{ind}} &=& \delta_{s,t}\delta_{ij}(1-m_i^2(t)) \\
\frac{\partial m_i(t)}{\partial J_{jk}}|_{\hbox{ind}} &=& \delta_{ik}(1-m_i^2(t))m_j(t-1) \\
\frac{\partial^2 m_i(t)}{\partial \theta_j(s)\partial \theta_k(s')}|_{\hbox{ind}} &=&  -2m_i(t)(1-m^2_i(t))\delta_{ij}\delta_{ik} \delta_{s,t} \delta_{s',t}\\
\frac{\partial^2 m_i(t)}{\partial J_{jk} \partial \theta_l(s)}|_{\hbox{ind}} &=&  
      -2m_i(t)(1-m^2_i(t))m_j(t-1)\delta_{ik}\delta_{il}\delta_{s,t} \nonumber \\
     &&  +(1-m^2_i(t))(1-m_j^2(t-1)) \delta_{ik}\delta_{kl}\delta_{s,t-1}\\
\frac{\partial^2 m_i(t)}{\partial J_{jk} \partial J_{lm}}|_{\hbox{ind}} &=& 
       \delta_{ik}(1-m_i^2(t)) (1-m_j^2(t-1))\delta_{lk} m_l(t-2) + (jk)\leftrightarrow (lm) \nonumber \\
    &&-2m_i(t)\delta_{ik}(1-m_i^2(t)) \delta_{im} \left(m_k(t-1)m_m(t-1) +\delta_{km}(1-m_k^2(t-1))\right)
\end{eqnarray}
To first order in $\epsilon$ (\ref{eq:Kappen-Spanjers-3.2}) hence gives
\begin{equation}
\sum_{s,j} \delta_{s,t}\delta_{ij}(1-m_i^2(t)) \Delta^{(1)}_j(s) + \sum_{jk} \delta_{ij}(1-m_i^2(t))m_k(t-1) J_{jk} = 0
\end{equation}
which is simply
\begin{equation}
A^{(1)}_i(t) \equiv \Delta^{(1)}_i(t) + \sum_{j} J_{ji}m_j(t-1)  = 0
\label{eq:A-first-order}
\end{equation} 
This is the same as "dynamic naive mean field"
\begin{equation}
\tanh^{-1}(m_i(t)) = \theta_i(t)+ \sum_{k} J_{ki} m_k(t-1)+  {\cal O}(\epsilon^2)
\end{equation}
The terms arising from second order derivatives and first order increments
can be grouped together as 
\begin{equation}
(1-m_i^2(t))\left(-m_i(t)(A^{(1)}_i)^2(t) - \sum_j(1-m_j^2(t-1))J_{ji}A^{(1)}_j(t-1)- m_i(t)\sum_k J_{ki}^2 (1-m_k^2(t-1))\right)
\label{eq:Kappen-Spanjers-A2}                         
\end{equation}
which together with the first order conditions (\ref{eq:A-first-order})
and the term from the first order derivative and second order increment
$(1-m_i^2(t))\Delta^{(2)}_i(t)$ gives
\begin{equation}
\Delta^{(2)}_i(t) = m_i(t)\sum_k J_{ki}^2 (1-m_k^2(t-1))
\label{eq:Kappen-Spanjers-A2-bis}                         
\end{equation}
This is the same as ``dynamic TAP'', compare (\ref{eq:tap_dynamic}) above
\begin{equation}
\tanh^{-1}(m_i(t)) = \theta_i(t)+ \sum_{k} J_{ki} m_k(t-1) -m_i(t)\sum_k J_{ki}^2 m_k^2(t-1)+  {\cal O}(\epsilon^3)
\end{equation} 

\section{The Information Geometry calculation to third order}
Third order contributions consist partly of terms involving lower than third order
derivatives and higher than first order increments. The calculation of these
use the same elements as above and are
\begin{eqnarray}
\sum_{j,s}\frac{\partial m_i(t)}{\partial \theta_j(s)}|_{\hbox{ind}} \Delta^{(3)}_j(s) &=& (1-m_i^2(t)) \Delta^{(3)}_i(t) \\
\sum_{j,s,k,s'}\frac{\partial^2 m_i(t)}{\partial \theta_j(s)\partial \theta_k(s')}|_{\hbox{ind}} \Delta^{(2)}_j(s) \Delta^{(1)}_k(s')&=&  -2m_i(t)(1-m^2_i(t))\Delta^{(2)}_i(t) \Delta^{(1)}_i(t)
\\
\sum_{j,k,l,s}\frac{\partial^2 m_i(t)}{\partial J_{jk} \partial \theta_l(s)}|_{\hbox{ind}} J_{jk} \Delta^{(2)}_l(s)
&=&  -2m_i(t)(1-m^2_i(t))\sum_j m_j(t-1)J_{ji} \Delta^{(2)}_i(t)\nonumber \\
&&  +(1-m^2_i(t))\sum_k(1-m_k^2(t-1)) J_{ki} \Delta^{(2)}_k(t-1) 
\end{eqnarray}
where two terms can be combined to
\begin{equation}
-2m_i(t)(1-m_i^2(t))\Delta^{(2)}_i(t)\left(\Delta^{(1)}_i(t)+\sum_k J_{ki}m_k(t-1)\right) = 0.
\end{equation}
The remainder is
\begin{equation}
 (1-m_i^2(t))\left(\Delta^{(3)}_i(t) + \sum_k (1-m_k^2(t-1)) J_{ki} \Delta^{(2)}_k(t-1)\right) \qquad\hbox{(lower order terms)} 
\end{equation}

To proceed with the terms from third order derivatives and first order increments it is useful
to introduce the streamlined notation
\begin{equation}
m_i = m_i(t)\qquad m_i'=m_i(t-1)\qquad m_i''=m_i(t-2)\qquad\hbox{etc.}
\end{equation}
and similar for all other quantities. It is also useful to note that though the derivatives
act on the complete expression involving both probability density $P$ and the $\tanh$
they partially obey a chain rule when taken to act on the magnetizations alone:
\begin{itemize}
\item a derivative with respect to an external field $\theta_j(s)$ functions as an ordinary
derivative and obeys a chain rule;
\item a derivative with respect to an interaction coefficient $J_{kl}$ acting on a once
or more than once primed quantity, such as $m_i'$ and $m_i''$, functions as an ordinary
derivative and obeys the chain rule;
\item a derivative with respect to an interaction coefficient $J_{kl}$ acting on
an unprimed quantity such as $m_i$ must be treated apart, since this derivative will
include a term taken on the $\tanh$, which in turn will give a higher order correlation.
\end{itemize}
These rules allow us to continue from what has already been computed and write
\begin{eqnarray}
\frac{\partial m_i}{\partial \theta_j(s)}|_{\hbox{ind}} &=& (1-m_i^2) \delta_{ij}\delta_{st} \nonumber \\
\frac{\partial^2 m_i}{\partial \theta_j(s)\partial \theta_k(s')}|_{\hbox{ind}} &=& 
-2m_i(1-m_i^2)\delta_{ij}\delta_{st}\delta_{ik}\delta_{s't} \nonumber \\
\frac{\partial^3 m_i}{\partial \theta_j(s)\partial \theta_k(s')  \partial \theta_l(s'')}|_{\hbox{ind}} &=& 
2(1-m_i^2)(3m_i^2-1) \delta_{ij}\delta_{st}\delta_{ik}\delta_{s't}\delta_{il}\delta_{s''t} \nonumber \\
&\vdots&\nonumber
\end{eqnarray}
For the mixed terms we have similarly
\begin{eqnarray}
\frac{\partial m_i}{\partial J_{jk}}|_{\hbox{ind}} &=& (1-m_i^2)\delta_{ik}m'_j \nonumber \\
\frac{\partial^2 m_i}{\partial J_{jk}\partial \theta_l(s)\partial }|_{\hbox{ind}} &=& 
-2m_i(1-m_i^2) \delta_{il}\delta_{s,t}\delta_{ik}m'_j + (1-m_i^2) \delta_{ik}(1-(m'_j)^2)\delta_{jl}\delta_{s,t-1} \nonumber \\
\frac{\partial^3 m_i}{\partial J_{jk}\partial \theta_l(s)\partial \theta_{l'}(s')}|_{\hbox{ind}} &=& 
2(1-m_i^2)(3m_i^2-1) \delta_{il'}\delta_{s',t}\delta_{il}\delta_{s,t}\delta_{ik}m'_j \nonumber \\
&-&2m_i(1-m_i^2) \delta_{il}\delta_{s,t}\delta_{ik}(1-(m'_j)^2)\delta_{l'j}\delta_{s',t-1} \nonumber \\
&-&2m_i(1-m_i^2)\delta_{il'}\delta_{s',t} \delta_{ik}(1-(m'_j)^2)\delta_{jl}\delta_{s,t-1} \nonumber \\
&+&(1-m_i^2) \delta_{ik}(-2m'_j)(1-(m'_j)^2)\delta_{jl}\delta_{s,t-1}\delta_{jl'}\delta_{s',t-1} \nonumber\\
&\vdots&\nonumber
\end{eqnarray}
and
\begin{eqnarray}
\frac{\partial^2 m_i}{\partial J_{jk} \partial J_{lm}}|_{\hbox{ind}} &=& 
       \delta_{ik}(1-m_i^2) (1-(m'_j)^2)\delta_{lk} m^{''}_l + (jk)\leftrightarrow (lm) \nonumber \\
    &&-2m_i(1-m_i^2)\delta_{ik} \delta_{im} \left(m'_km'_m+\chi'_{km})\right)\nonumber \\
\frac{\partial^3 m_i}{\partial J_{jk} \partial J_{lm} \partial \theta_n(s)}|_{\hbox{ind}} &=& 
       \delta_{ik}(-2m_i(1-m_i^2)\delta_{in}\delta_{s,t} (1-(m'_j)^2)\delta_{lk} m^{''}_l\nonumber \\
       &&\quad + \delta_{ik}(1-m_i^2)(-2m'_j(1-(m'_j)^2)\delta_{jn}\delta_{s,t-1}\delta_{lk} m^{''}_l\nonumber \\
       &&\quad\quad +\delta_{ik}(1-m_i^2)(1-(m'_j)^2)\delta_{lk} (1-(m^{''}_l)^2)\delta_{ln}\delta_{s,t-2}\nonumber \\
     &+& (jk)\leftrightarrow (lm) \nonumber \\
    &+&2(1-m_i^2)(3m_i²-1)\delta_{in}\delta_{s,t}\delta_{ik} \delta_{im} <\sigma_k(t-1)\sigma_m(t-1)>\nonumber \\
    &&\quad -2m_i(1-m_i^2)\delta_{ik} \delta_{im} (1-(m'_k)^2)\delta_{kn}\delta_{s,t-1} m'_m\nonumber \\
    &&\quad\quad -2m_i(1-m_i^2)\delta_{ik} \delta_{im} m'_k (1-(m'_m)^2)\delta_{mn}\delta_{s,t-1}\nonumber \\
    &&\quad\quad\quad -2m_i(1-m_i^2)\delta_{ik} \delta_{im} \frac{\partial \chi'_{km}}{\partial\theta_n(s)}  \nonumber \\
&\vdots& \nonumber
\end{eqnarray}
where we use the correlation function $\chi_{km}=<\sigma_k(t)\sigma_m(t)>-m_k m_m$.
Its partial derivative with respect to an external field is always zero, and
the last term in above therefore vanishes.
The more cumbersome term is three derivatives with respect to interaction
coefficients, which we can start from
\begin{equation}
\frac{\partial^3 m_i}{\partial_{pq}\partial J_{jk} \partial J_{lm}}|_{\hbox{ind}} =
\frac{\partial}{\partial J_{pq}}\left[\begin{array}{l}
       \sum_{\sigma}\frac{\partial^2 P(\sigma)}{\partial J_{jk} \partial J_{lm}}\tanh(\cdot) +\\
       \sum_{\sigma}\frac{\partial P(\sigma)}{\partial J_{jk}}(1-\tanh^2(\cdot))\delta_{im}\sigma_l(t-1)\\
       \qquad  + (jk)\leftrightarrow (lm)\\
       \sum_{\sigma}P(\sigma)(-2\tanh(\cdot))(1-\tanh^2(\cdot))\delta_{im}\sigma_l(t-1)\delta_{ik}\sigma_j(t-1)\\
       \end{array}\right]
\end{equation}
Applying $\partial J_{pq}$ gives (at least conceptually) eight terms. The term from acting on
$\frac{\partial^2 P(\sigma)}{\partial J_{jk} \partial J_{lm}}$ vanishes. The term from acting on
$\tanh(\cdot)$ in the first line gives a second derivative with respect to interaction coefficients
of a magnetization. The terms from the second and the third line give combinations
involving either second derivatives of a magnetization, or first derivatives
of a correlation function. The terms from the last line are a third order correlation
function and further first deritives of second order correlation functions.

Taking all together we can sum the contributions to
\begin{eqnarray}\label{eq:horrible-things}
\hbox{Third order} &=& \frac{1}{6}2(1-m_i^2)(3m_i^2-1)(A^{(1)}_i(t))^3 \nonumber \\
                   &+& 2m_i(1-m_i^2)\Delta^{(1)}_i(t)\sum_l (1-(m'_l)^2)J_{li}A^{(1)}_l(t-1)\nonumber \\
                   &-&(1-m_i^2)\sum_l J_{li}(A^{(1)}_l(t))^2 \nonumber \\
                   &+&\frac{1}{2}2(1-m_i^2)(3m_i^2-1)A^{(1)}_i(t)\sum_{lm}J_{il}J_{im}\chi_{lm} \nonumber \\
                   &-&2m_i(1-m_i^2)\sum_{ln} J_{li}(1-(m'_l)^2)A^{(1)}_l(t-1)J_{ni}m'_{n} \nonumber \\
                   &+&(1-m_i^2)\sum_{ml}J_{mi}J_{lm}(1-(m'_m)^2)(1-(m^{''}_l)^2)A^{(1)}_l(t-2)  + (m)\leftrightarrow (l)\nonumber \\
                   &-&m_i(1-m_i^2)\sum_{ln, js} J_{li} J_{ni}\left(\frac{\partial \chi_{ln}(t-1)}{\partial \theta_j(s)}\right) \Delta^{(1)}_j(s) \nonumber \\
                   &-&\frac{1}{3}m_i(1-m_i^2)\sum_{ln, js} J_{li} J_{ni}\left(\frac{\partial \chi_{ln}(t-1)}{\partial J_{pq}}\right) J_{pq} + \hbox{circ. perm.} \nonumber \\
                   &+&\frac{1}{3}(1-m_i^2)(3m_i^2-1)\sum_{lnq}J_{li}J_{ni}J_{qi}\chi'_{lnq} 
\end{eqnarray} 
where in the last line we have used $\chi_{lnq}=<(\sigma_l(t)-m_l)(\sigma_n(t)-m_n)(\sigma_q(t)-m_q)>$.
All the terms in above containing the first order terms $A^{(1)}$ vanish, the partial
derivative terms of the second order correlation function with respect to external field vanish, and the last
line is at least smaller than $\epsilon^3$. The sole remaining terms hence come from
the partial derivatives of second order correlation functions with respect to
interaction parameters. These are model dependent, and are evaluated to
non-zero for the \textit{sequential update} rule in~\cite{KappenSpanjers2000}.
For the \textit{parallel update} rule which we look at here they are however
zero. The collection of terms (\ref{eq:horrible-things}) therefore evaluates to zero.   


\end{document}